\documentclass[pra,twocolumn,amsmath,amssymb,floatfix,reprint,footinbib,superscriptaddress,longbibliography,showkeys]{revtex4-2}
\usepackage{picinpar,graphicx}

\usepackage{braket}
\usepackage{amsmath}
\usepackage{graphicx}
\usepackage{epstopdf}
\usepackage{dcolumn}
\usepackage{graphicx}
\usepackage{mathrsfs}
\usepackage{mdwlist}
\usepackage{subfigure}
\usepackage{booktabs}
\usepackage{amsmath}
\usepackage{dsfont}
\usepackage{amstext}
\usepackage{amssymb}
\usepackage{amsbsy}
\usepackage{bbm}
\usepackage{amsthm}
\usepackage{graphicx}
\usepackage{textcomp}
\usepackage{color}
\usepackage[colorlinks,citecolor=blue]{hyperref}
\usepackage{lipsum}
\definecolor{Dgreen}{RGB}{0, 100, 0}
\usepackage{url}
\usepackage[colorlinks]{hyperref}
\hypersetup{%
	plainpages=true,
	breaklinks=true,  
	hypertexnames=false,  
	pageanchor=true,
	colorlinks=true,
	linkcolor={blue},
	citecolor={red},
	urlcolor={blue},
	anchorcolor={black}
}

\begin{document}
	
	\title{Giant-Atom Quantum Batteries}
	\author{Ke-Xiong Yan}
	\affiliation{Fujian Key Laboratory of Quantum Information and Quantum Optics, Fuzhou University, Fuzhou 350116, China}
	\affiliation{Department of Physics, Fuzhou University, Fuzhou, 350116, China}
        \affiliation{Quantum Information Physics Theory Research Team, Center for Quantum Computing, RIKEN, Wako-shi, Saitama 351-0198, Japan}

	\author{Yang Liu}
	\affiliation{Center for Joint Quantum Studies and Department of Physics, School of Science, Tianjin University, Tianjin 300350, China}
	
	\author{Yang Xiao}
	\affiliation{Fujian Key Laboratory of Quantum Information and Quantum Optics, Fuzhou University, Fuzhou 350116, China}
	\affiliation{Department of Physics, Fuzhou University, Fuzhou, 350116, China}
	
	\author{Jun-Hao Lin}
	\affiliation{Fujian Key Laboratory of Quantum Information and Quantum Optics, Fuzhou University, Fuzhou 350116, China}
	\affiliation{Department of Physics, Fuzhou University, Fuzhou, 350116, China}

	\author{\\Jie Song}
	\affiliation{Department of Physics, Harbin Institute of Technology, Harbin 150001, China}

	\author{Ye-Hong Chen}\thanks{yehong.chen@fzu.edu.cn}
	\affiliation{Fujian Key Laboratory of Quantum Information and Quantum Optics, Fuzhou University, Fuzhou 350116, China}
	\affiliation{Department of Physics, Fuzhou University, Fuzhou, 350116, China}
	\affiliation{Quantum Information Physics Theory Research Team, Center for Quantum Computing, RIKEN, Wako-shi, Saitama 351-0198, Japan}%
	
	\author{Franco Nori}
	\affiliation{Quantum Information Physics Theory Research Team, Center for Quantum Computing, RIKEN, Wako-shi, Saitama 351-0198, Japan}%
	\affiliation{Department of Physics, University of Michigan, Ann Arbor, Michigan 48109-1040, USA}
	
	\author{Yan Xia}\thanks{xia-208@163.com}
	\affiliation{Fujian Key Laboratory of Quantum Information and Quantum Optics, Fuzhou University, Fuzhou 350116, China}
	\affiliation{Department of Physics, Fuzhou University, Fuzhou, 350116, China}

	
	\date{\today}
	\begin{abstract}
		Environmentally induced decoherence poses a fundamental challenge to quantum energy storage systems, causing irreversible energy dissipation and performance aging of quantum batteries (QBs). To address this issue, we propose a QB protocol utilizing the nonlocal coupling properties of giant atoms (GAs). In this architecture, both the QB and its charger are implemented as superconducting GAs with multiple nonlocal coupling points to a shared microwave waveguide. By engineering these atoms in a braided configuration—where their coupling paths are spatially interleaved—we show the emergence of decoherence-immune interaction dynamics. This unique geometry enables destructive interference between decoherence channels while preserving coherent energy transfer between the charger and the QB, thereby effectively suppressing the aging effects induced by waveguide-mediated dissipation. The charging properties of separated and nested coupled configurations are investigated. The results show that these two configurations underperform the braided configuration. Additionally, we propose a long-range chiral charging scheme that facilitates unidirectional energy transfer between the charger and the battery, with the capability to reverse the flow direction by modulating the applied magnetic flux. Our result provides guidelines for implementing a decoherence-resistant charging protocol and remote chiral QBs in circuits with GAs engineering. 
	\end{abstract}
	
	\maketitle

	\textit{Introduction.}---As an energy storage device governed by quantum mechanical principles, quantum batteries (QBs) have emerged as one of the most promising applications related to future quantum technologies~\cite{RevModPhys.96.031001,d9k1-75d4,PhysRevLett.122.210601,PRXQuantum.5.030319,PhysRevLett.132.210402,PhysRevLett.134.180401,PhysRevResearch.2.023095,RevModPhys.96.031001}. Distinct from classical counterparts, QBs fundamentally exploit quantum resources—such as quantum coherence and multipartite entanglement—to enable superior energy storage and transmission characteristics~\cite{PhysRevLett.111.240401,Binder2015,PhysRevLett.122.047702}. Theoretical investigations demonstrate that these non-classical correlations can enhance key performance metrics: (i) exponentially accelerated charging power through collective quantum effects~\cite{PhysRevLett.120.117702,PhysRevB.102.245407,PhysRevB.102.245407,Chen2020,PhysRevA.103.052220,PhysRevResearch.2.023113,Jose2024,PhysRevLett.118.150601}, (ii) superextensive scaling of storage capacity with system size~\cite{WangYI2025,WangYA2025,PhysRevLett.131.030402}, and (iii) nonequilibrium work extraction surpassing classical counterparts~\cite{PhysRevLett.125.180603,PhysRevLett.129.130602,PhysRevE.102.042111}.
	
	While QBs exhibit superior performance compared to classical counterparts, their stored energy becomes compromised through environmentally induced decoherence during prolonged storage, a phenomenon known as the aging of the QB~\cite{PhysRevA.100.043833}. This inherent limitation necessitates the development of decoherence suppression strategies for practical quantum energy device implementation. Two complementary anti-decoherence frameworks  have been proposed: (1) \textit{intrinsic protection protocols} leveraging symmetry-protected decoherence-free subspaces~\cite{LSH2019}, destructive interference-enabled dark-state engineering~\cite{PhysRevApplied.14.024092}, and periodic-driving-modulated Floquet-engineered interactions~\cite{PhysRevA.102.060201}; and (2) \textit{extrinsic environmental control} encompassing real-time quantum feedback control with measurement-based error correction~\cite{Mitchison2021}, environment engineering~\cite{PhysRevA.106.012425,PhysRevE.105.064119,PhysRevE.104.064143}, and sequential measurements~\cite{PhysRevResearch.2.013095}. However, the simultaneous suppression of decoherence-induced energy loss and realization of high-efficiency charging in QBs remains an open question.
	
	In this manuscript, we propose a QB protocol that leverages the non-local coupling properties of giant atoms (GAs) to address the challenges of \textit{aging} and \textit{inefficiency}. As artificial atoms, GAs can achieve coupling to the boson field at points separated by a few wavelengths~\cite{PhysRevA.90.013837}. This feature has led to extensive research~\cite{PhysRevResearch.6.013301,PhysRevResearch.6.013279,PhysRevResearch.6.033243,PhysRevResearch.6.043222,PhysRevResearch.5.043135,PhysRevResearch.4.023198,PhysRevResearch.2.043184,Gu2017,Kockum2019}. For a GA with multiple coupling points, the relaxation rate and the Lamb shift can be controlled by tuning the atomic transition frequency~\cite{PhysRevA.90.013837,PhysRevA.106.063717}. In the case of multiple GAs with multiple coupling points, the interaction strengths, as well as the individual and collective relaxation rates, can be controlled by varying the coupling strength of each coupling point as well as the distance between the coupling points~\cite{PhysRevLett.120.140404,Kannan2020}. Additionally, interference effects resulting from the non-local coupling of giant atoms give rise to chiral bound states, enabling unidirectional energy transfer between the charger and the battery~\cite{PhysRevLett.126.043602,Wang2022}. Furthermore, by modulating the atom-waveguide coupling, we can tune the chirality, thereby offering flexible control over the energy storage and extraction modes in the QB.

	In our proposal, two GAs (each coupled to a waveguide at two spatially separated points) are configured as the charger and battery element, respectively. We study the charging property of three different coupling configurations: braided, separated, and nested couplings~\cite{PhysRevLett.120.140404}. 
    
    When the two GAs are in a braided configuration, by engineering the distance between connection points, we can \textit{observe persistent energy transfer between the charger and battery without direct coupling}. 
    
    Simultaneously, \textit{all} dissipation channels are eliminated due to their mutual interference, thereby enabling the battery to \textit{overcome the challenges associated with aging and inefficiency}. 
    
    For the separated and nested structures, we find that the GAs of both configurations underperform the braided configuration in terms of charging performance due to the presence of dissipation in the charging process. 
    
    Furthermore, we investigate the possibility of separated GAs to enable \textit{chiral charging over extended distances}. Simulation results show that energy can flow unidirectionally between the charger and the battery without loss.

	\begin{figure}
		\centering
		\includegraphics[scale=0.17]{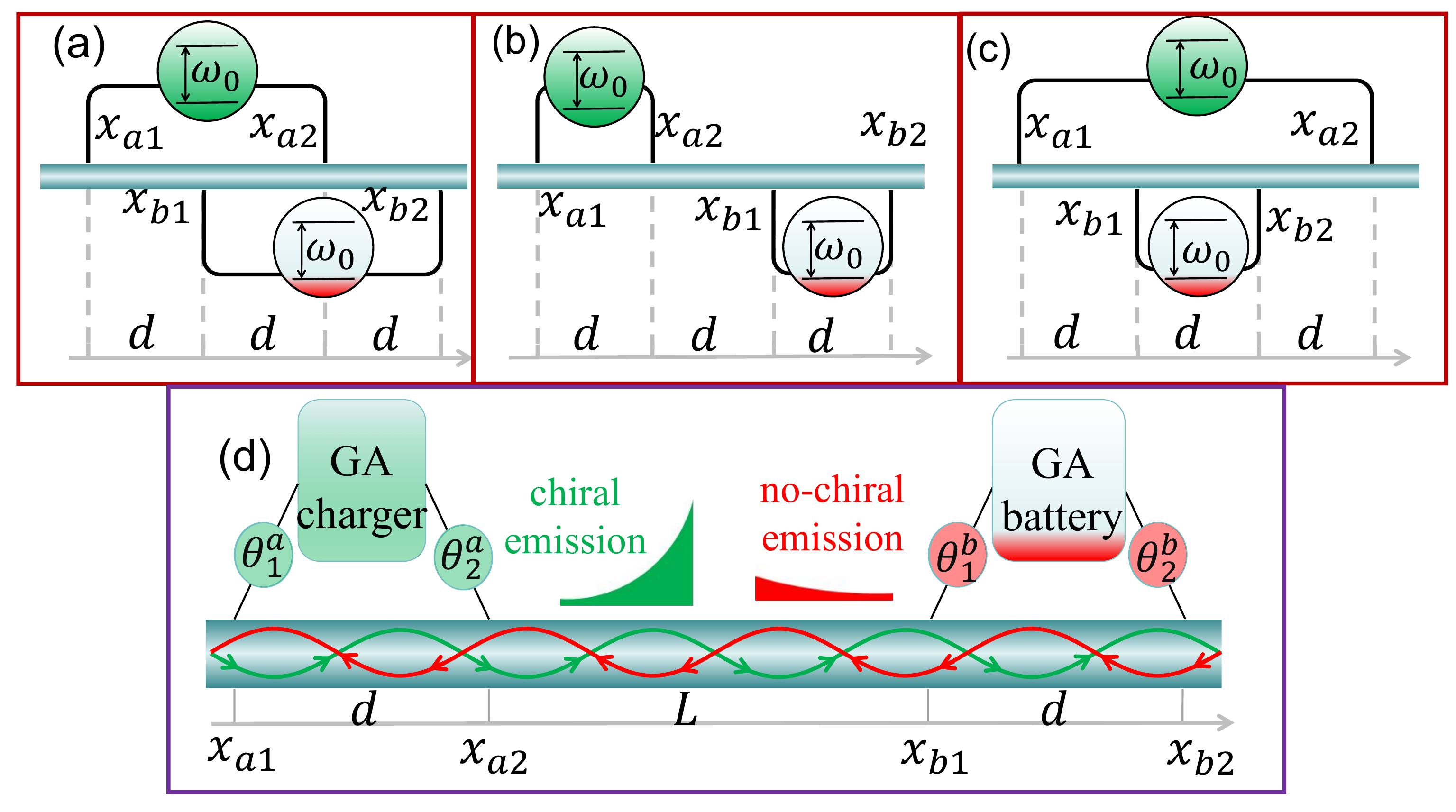}
		\caption{Schematics of the (a) braided; (b) separated; (c) nested configurations for double two-level giant atoms with energy separation $\omega_{0}$ interacting with a common waveguide. (d) Schematic diagram of our remote chiral charging protocol. The phases $\theta_{1,2}^{a(b)}$ are modulated by the external magnetic flux.}
		\label{F1}
	\end{figure}
	\textit{Model.}---The charging QB protocol consists of two GAs acting as the charger and the QB, in which each GA interacts with a common 1D waveguide through two separated coupling points, as shown in Fig.~\ref{F1}. The location of these coupling points are labeled by the coordinates $x_{jn}$, with $j=a, b$ marking the GAs and $n=1, 2$ denoting the two coupling points of each atom. Under the rotating-wave approximation, the system Hamiltonian reads (hereafter we set $\hbar=1$)
    \begin{eqnarray}
    	{H} &=& \omega_0 \sum_{j=a,b} {\sigma}_j^+ {\sigma}_j^- + \sum_{k=-\infty}^{+\infty} \big[ \omega_k{a}_k^\dagger {a}_k  \\\nonumber
    	&& + \sum_{j=a,b} \sum_{n=1,2} \left( g_{jn} {\sigma}_j^+ {a}_k + \text{H.c.} \right) \big],
    \end{eqnarray}
    where ${\sigma}_j^-=\ket{g_j}\bra{e_j}$ is the transition operator from the excited state $\ket{e_j}$ to the ground state $\ket{g_j}$ with frequency $\omega_{0}$ of the charger when $j=a$ and of the QB when $j=b$ and ${a}_k$ is the annihilation operator of the $k$th mode with frequency $\omega_k$ in the waveguide. The constant term $g_{jn}$ is the coupling strength related to the coupling points $x_{jn}$. 
    
    Note that we apply the Wigner-Weisskopf approximation~\cite{walls2008optics,Agarwal2012,Marlan1997} in the weak-coupling regime, so the coupling strength can be treated as independent of $k$. For simplicity, we assume that the coupling strength at each connection point is $g$ (i.e. $g_{jn}\equiv g$) and the distance between neighboring coupling points is $d$~(see Fig.~\ref{F1}).

    \textit{Charging performance.}---To study the charging performance of the two GAs, we treat the fields in the waveguide as the environment for the two GAs~\cite{PhysRevA.108.023728}. The evolution of the two atoms is governed by the quantum master equation.  Assuming $\omega_k \approx \omega_0+(k-k_{0})v_{g}$, with $k_{0}$ ($v_{g}$) the wave number (group velocity) of the field at frequency $\omega_{0}$~\cite{PhysRevLett.95.213001,PhysRevA.79.023837}, the dynamics of the two GAs in these three different coupling configurations are governed by the unified master equation~\cite{PhysRevA.108.023728}
    \begin{eqnarray}
    	\label{eq2}
    	\dot{{\rho}}_I&=&-i[{H}_{\rm eff}, {\rho}_I]+\sum_{j=a,b} \Gamma_{j}D[{\sigma}_{j}^{-}]\\ \nonumber
    	&&+\Gamma_{\rm coll}\{D[{\sigma}_{a}^{-},{\sigma}_{b}^{+}]+{\rm H.c.}\},
    \end{eqnarray}   
   where $D[A]=A{\rho}_IA^\dagger-1/2(A^\dagger A{\rho}_I+{\rho}_IA^\dagger A)$, $D[A,B]=A{\rho}_IB^\dagger-1/2(A^\dagger B{\rho}_I+{\rho}_IA^\dagger B)$, $\Gamma_{j=a,b}=\sum_{n,m=1,2}\gamma \cos(k_{0}|x_{jn}-x_{jm}|)$ and $\Gamma_{\rm coll}=\gamma \cos(k_{0}|x_{an}-x_{bm}|)$ are the individual and collective decay rates of the GAs. The parameter $\gamma=4\pi g^2/v_{g}$ is the bare relaxation rate at each connection point.
   
   The effective Hamiltonian ${H}_{\rm eff}$ of the two GAs in Eq.~({\ref{eq2}}) takes the form
   \begin{eqnarray}
   	\label{eq3}
   	{H}_{\rm eff}=\sum_{j=a,b}\delta \omega_{j}{\sigma}_j^+ {\sigma}_j^- +\sum_{i\neq j}g_{ab}{\sigma}_i^+ {\sigma}_j^-.
   \end{eqnarray}
    where $\delta \omega_{j}=\sum_{n,m=1,2}\gamma \sin(k_0|x_{jn}-x_{jm}|)/2$ is the Lamb shift of the $j$th GA and $g_{ab}=\sum_{n,m=1,2}\gamma \sin(k_0|x_{an}-x_{bm}|)/2$ is the exchange interaction strength. Note that Eq.~(\ref{eq2}) is consistent with the quantum master equation derived by the SLH formalism in Ref.~\cite{PhysRevLett.120.140404}. The Hamiltonian ${H}_{\rm eff}$ in Eq.~(\ref{eq3}) reveals that, \textit{although a direct charger-QB interaction is absent, an effective charger-QB coupling is induced by the mediation role of the electromagnetic fields in the waveguide}.

    \textit{QB performance indicators.}---In the process of charging, the charger is intialized to its excited state by a pump laser and then the QB is continuously energised from the charger~\cite{RevModPhys.96.031001,PhysRevLett.132.090401}. The energy of the QB is 
	\begin{equation}
		E(t)=\operatorname{Tr}[{\rho}_{B}(t){H}_{B}],
	\end{equation}
	where ${H}_{B}=\omega_{0}{\sigma}_{b}^{+}{\sigma}_{b}^{-}$ and ${\rho}_{B}(t)=\operatorname{Tr}_{a}[{\rho}_{I}]$. The maximal amount of work that can be extracted form a state ${\rho}_{B}(t)$ is provided by the ergotropy~\cite{PhysRevLett.125.180603,PhysRevE.102.042111,RevModPhys.96.031001}:
	\begin{equation}
		\mathcal{E}(t)=\operatorname{Tr}[{\rho}_{B}(t){H}_{B}]-\operatorname{Tr}[\tilde{{\rho}}_{B}(t){H}_{B}],
	\end{equation}
	where ${\tilde{\rho}}_{B}(t)=\sum_{m}r_{m}(t)\ket{s_m}\bra{s_m}$ is the passive state, $\{r_{m}(t)\}$ are the eigenvalues of ${\rho}_{B}(t)$ ordered in a descending sort, and $\{\ket{s_m}\}$ are the eigenstates of ${H}_B$ with the corresponding eigenvalues $s_m$ ordered in an ascending sort. The fluctuation between the intial and final time of the charging process is represented by the correlator~\cite{Friis2018,PhysRevB.102.245407}
	\begin{equation}
		\Sigma(t)=\Big[\sqrt{\left\langle H_{B}^{2}(t)\right\rangle-\left\langle H_{B}(t)\right\rangle^{2}}-\sqrt{\left\langle H_{B}^{2}(0)\right\rangle-\left\langle H_{B}(0)\right\rangle^{2}}\Big],
	\end{equation}
	where $H_{B}(t)$ is the Heisenberg time evolution of the operator $H_{B}$. The average charging power of QB is given by~\cite{RevModPhys.96.031001}
	\begin{equation}
		\mathcal{P}(t)=\frac{\mathcal{E}(t)}{t}.
	\end{equation} We investigate the charging property of the two GAs in the three different coupling configurations shown in Fig.~\ref{F1}. For each configuration, the initially state of the GA system is $\ket{\psi(0)}=\ket{e_a}\ket{g_b}$.

    \textit{Charging characteristics of braided GAs.}---In the braided configuration, the two inner connection points are placed in between the two connection points of the other atoms, as shown in Fig.~\ref{F1}{\color{blue}(a)}. We can obtain the Lamb shifts $\delta \omega_{a}= \delta \omega_{b}=\gamma \sin2\theta$, the exchange coupling strength $g_{ab}=\gamma(3\sin \theta+\sin 3\theta)/2$, the individual decay rates $\Gamma_{a}=\Gamma_{b}=2\gamma(1+\cos 2\theta)$, and the collective decay rate $\Gamma_{\rm coll}=\gamma(3\cos \theta +\cos 3\theta)$~(see the Supplemental Material~\cite{YKX250527} for details). The parameter $\theta=k_{0}d$ (where $k_{0}=\omega_{0}/v_{g}$) denotes the phase accumulated by the electromagnetic field as it propagates through the waveguide between two neighboring connection points. By substituting the expressions of $g_{ab}$, $\Gamma_{a}$, $\Gamma_{b}$, and $\Gamma_{\rm coll}$ into Eq.~(\ref{eq2}), the charging property of the two separated GAs can be analyzed.
	
	\begin{figure*}
		\centering
		\includegraphics[scale=0.13]{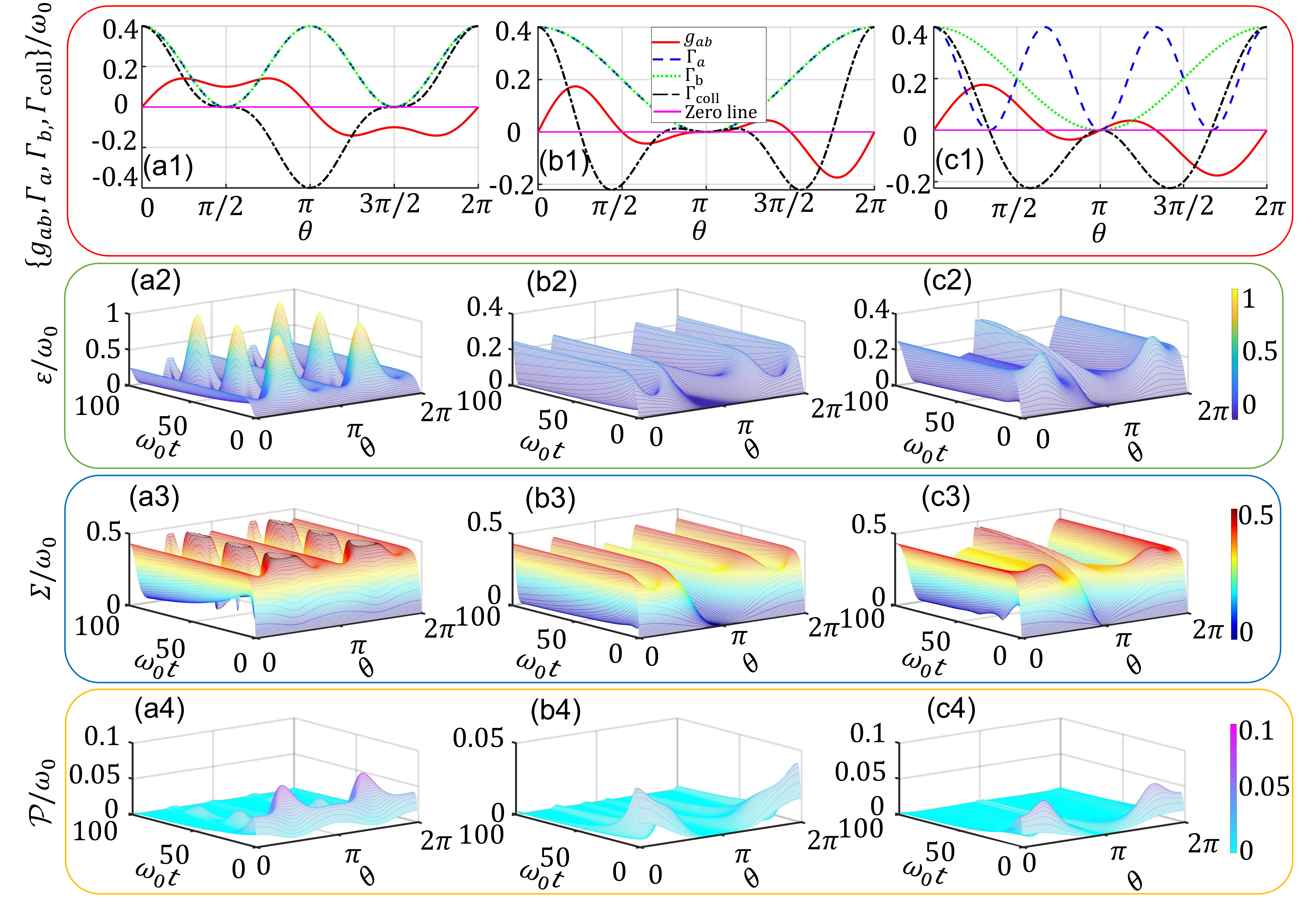}
		\caption{(a1)-(c1) Exchange interaction $g_{ab}$ (red solid curve), individual decay rate $\Gamma_{a}$ (blue dashed curve), individual decay rate $\Gamma_{b}$ (green dotted curve), and collective decay rate $\Gamma_{\rm coll}$ (black dash-dot curve) as a function of $\theta$ for the braided (a1), separated (b1), and nested (c1) coupling configuration. (a2)-(c2) Ergotropy $\mathcal{E}$, (a3)-(c3) Fluctuations $\Sigma$, and (a4)-(c4) Average charging power $\mathcal{P}$, as functions of the scaled charging time $\omega_{0}t$ and the accumulated phase $\theta$. The parameter $\gamma=0.1\omega_{0}$.}	
		\label{F7}
	\end{figure*}
	\begin{figure}
		\centering
		\includegraphics[scale=0.35]{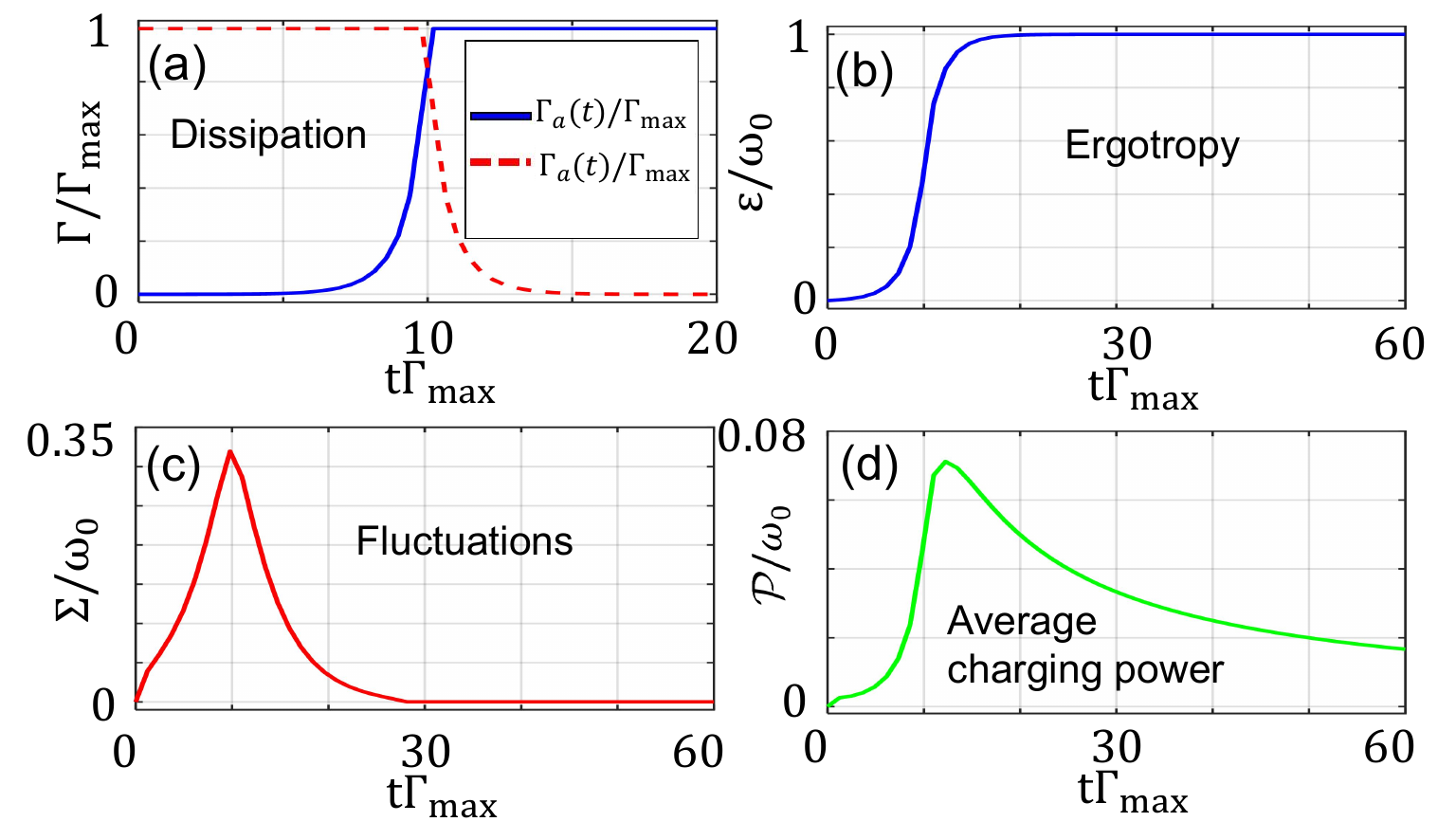}
		\caption{(a) Dissipation of the two separated GAs with the propagation time $\tau = 10\Gamma_{\rm max}$. (b) Ergotropy $\mathcal{E}$, (c) Fluctuations $\Sigma$, and (d) Average charging power $\mathcal{P}$, as functions of the scaled charging time $\Gamma_{\rm max}t$. }	
		\label{F8}
	\end{figure}

	In Fig.~\ref{F7}{\color{blue}(a1)}, we plot the exchange interaction, the individual dissipation rate, and the collective dissipation rate of the two GAs as a function of the phase shift $\theta$. It can be seen that both the individual and collective dissipation rates are zero when the phase shift $\theta=(n+1/2)\pi$, with an integer $n$, while the exchange interaction is non-zero. This suggests that \textit{decoherence-immune energy transfer between the charger and the battery} can be achieved when the distance between neighboring connection points is $d=(1+2n)/4\lambda_{0}$ in the braided configuration. The parameter $\lambda_{0}=2\pi v_{g}/\omega_{0}$ is the wavelength of the electromagnetic wave in the waveguide with frequency $\omega_{0}$. 
	
	From Figs.~\ref{F7}{\color{blue}(a2)}, \ref{F7}{\color{blue}(a3)} and \ref{F7}{\color{blue}(a4)} we see that the ergotropy $\mathcal{E}$, the fluctuation $\Sigma$, and the average power $\mathcal{P}$ are modulated by the phase shift $\theta$. Meanwhile, the dependence of $\mathcal{E}$, $\Sigma$, and $\mathcal{P}$ on $\theta$ is a $\pi$-period function. In can be found from Fig.~\ref{F7}{\color{blue}(a2)} that there are two ridges exhibiting a Rabi-like oscillating process when $\theta$ is near either $\pi/2$ or $3\pi/2$, implying a lossless energy transfer between the charger and the battery.  
	
	In particular, when $\theta=n\pi$, $\mathcal{E}$ approaches a steady value $0.25\omega_{0}$ in the long-time limit. This is because when $\theta \to n\pi$, we obtain $|\Gamma_{{\rm coll},a,b}| \to 0.4\omega_{0}$ and $g_{ab} \to 0$, which leads to the emergence of a steady state and a nonoscillatory charging process.
	
    When $\theta = \pi/2$ or $\theta = 3\pi/2$, the energy fluctuation and charging power over time reflect the charging characteristics of the energy lossless charging process. During this charging process, the maximum average charging power is $0.072\omega_{0}$ and the maximum energy fluctuation is $0.5\omega_{0}$. This numerical result is consistent with the charging characteristic of the energy transfer process between two two-level atoms under a closed system~\cite{PhysRevB.98.205423}. When $\theta = n\pi$, the energy fluctuates as well as the average power does not oscillate over time. The maximum value of the average power and the maximum energy fluctuation during this charging process are $0.0407\omega_{0}$ and $0.4329\omega_{0}$, respectively.

    \textit{Charging characteristics of separated GAs.}---We now turn to the case of two separated GAs, as shown in Fig.~\ref{F1}{\color{blue}(b)}. In this configuration, we can obtain the Lamb shifts $\delta \omega_{a}=\delta \omega_{b}=\gamma \sin \theta$, the exchanging coupling strength $g_{ab}=\gamma(\sin \theta+2\sin 2\theta +\sin 3\theta )/2$, the individual decay rates $\Gamma_{a}=\Gamma_{b}=2\gamma(1+\cos \theta)$, and the collective decay rate $\Gamma_{\rm coll}=\gamma(\cos \theta + 2\cos 2\theta+ \cos 3\theta)$ (see the Supplemental Material~\cite{YKX250527} for details). From Fig.~\ref{F7}{\color{blue}(b1)}, it can be observed that when $\theta =\pi$, the interaction strength $g_{ab}$ and the decay rates $\Gamma_{{\rm coll},a,b}$ become zero, indicating that the two separated GAs are \textit{decoupled} from the waveguide. In the separated configuration, lossless charging of two GAs is \textit{not} possible regardless of the value of $\theta$. The maximum energy achievable for the GA $b$ as a battery is $0.250\omega_{0}$. The maximum energy fluctuation during charging and the maximum power are $0.443\omega_{0}$ and $0.040\omega_{0}$, respectively. 
    
    In Fig.~\ref{F7}{\color{blue}(b2)}, there are four ridges with heights exceeding $0.2\omega_{0}$. These ridges do not oscillate with time, indicating that the system is in a steady state. When the phase shift $\theta =\pi$, the ergotropy $\mathcal{E}$, fluctuation $\Sigma$, and power $\mathcal{P}$ are all \textit{zero}. This is caused by \textit{the inability of the charger and battery to produce indirect coupling through the waveguide}. We can also find that the emergence of a nonoscillatory charging process when $\theta \to 0,2\pi$, similar to the case of $\theta \to n\pi$ in the braided configuration. In Fig.~\ref{F7}{\color{blue}(b3)}, the variation of energy fluctuations with the phase shift is similar to that of ergotropy. This is due to the non-oscillatory nature of the charging process, and a similar phenomenon is observed in the nested configuration. From Fig.~\ref{F7}{\color{blue}(b4)}, we can see that the power reaches its maximum value only when $\theta=0$ and $\theta=\pi$, which is a feature unique to the separated configuration.

     \textit{Charging characteristics of nested GAs.}---Finally, we investigate the charging property for the nested coupling configuration, as depicted in Fig.~\ref{F1}{\color{blue}(c)}. In this case, the relevant parameters in the quantum master equation~(\ref{eq2}) are given by $\delta \omega_{a}=\gamma \sin 3\theta$, $\delta \omega_{b}=\gamma \sin \theta$, $g_{ab}=\gamma(\sin \theta +\sin 2\theta)$, $\Gamma_{a}=2\gamma(1+\cos3\theta)$, $\Gamma_{b}=2\gamma(1+\cos \theta)$, and $\Gamma_{\rm coll}=2\gamma(\cos\theta+\cos2\theta)$, respectively (see the Supplemental Material~\cite{YKX250527} for details).
    From Fig.~\ref{F7}{\color{blue}(c1)}, it can be found that the two GAs in the nested configuration are decoupled from the waveguide when $\theta=\pi$. When $\theta=2n\pi$, the indirect interaction of the two GAs is zero, but the collective and individual dissipation rates into the waveguide reach a maximum, which can lead to the emergence of a non-oscillatory charging process. 
    
    From Figs.~\ref{F7}{\color{blue}(c2)}, \ref{F7}{\color{blue}(c3)} and \ref{F7}{\color{blue}(c4)}, it can be found that the maximum values of the ergotropy, the energy fluctuation, and the average power are $0.329\omega_{0}$, $0.469\omega_{0}$, and $0.057\omega_{0}$, respectively.
    Both ergotropy $\mathcal{E}$ and fluctuations $\Sigma$ converge rapidly to zero when $\theta \to \pi$. The physical mechanism behind this phenomenon is that the excitation paths of the two GAs interfere \textit{destructively}, resulting in the absence of energy transfer channels between the two GAs and between the GAs and the waveguide. The same phenomenon occurs in both the nested and separated configurations, with the only difference being the rate at which the energy transfer channel disappears as the phase changes.

       \textit{Remote chiral charging using GAs.}---In a long-distance chiral charging protocol [see Fig.~\ref{F1}{\color{blue}(d)}], GAs are suitable in a separated configuration. The distance between the charger and the battery is defined as $L=x_{b1}-x_{a2}$. To achieve \textit{unidirectional energy transfer}, a phase difference must be introduced between the two coupling points of the GA. This can be accomplished by applying a magnetic flux $\Phi_{\rm ext}^{a(b)}$ that time-modulates each coupling point individually~\cite{PhysRevLett.101.100501,PhysRevA.81.042304,PhysRevA.80.062109,PhysRevA.78.063827,W2022}. We assume that the phase difference between the two coupling points of the GA $a$ and the GA $b$ are $\theta^{a}=\theta_2^{a}-\theta_1^{a}$ and $\theta^{b}=\theta_2^{b}-\theta_1^{b}$, respectively. When the phase relation $\theta-\theta^{a(b)}=(2N+1)\pi$, the GAs decouple with the left-passing modes due to the destructive interference. We can obtain the Lamb shifts $\delta \omega_{a}=\delta \omega_{b}=-\gamma_{a(b)}(t) \sin 2\theta$ and the exchange coupling strength $g_{ab}=i\sqrt{\gamma_{a}(t)\gamma_{b}(t)}\sin^2\theta$. The jump operator for the master equation in Eq.~(\ref{eq2}) contains only the right-passing mode, i.e. $\dot{{\rho}}_I=-i[{H}_{\rm eff},\label{eq4} {\rho}_I]+D[{L}_{R}]$, where ${L}_{R}=i (\sqrt{\Gamma_{a}(t)/2}{\sigma}_{a}^{-}+\sqrt{\Gamma_{b}(t)/2}{\sigma}_{b}^{-})$ and $\sqrt{\Gamma_{a(b)}(t)/2}=2\sin \theta \sqrt{\gamma_{a(b)}(t)/2}$ (see the Supplemental Material~\cite{YKX250527} for details).
    
    By experimentally varying the magnetic flux at the coupling points~\cite{Wulschner2016,PhysRevA.92.012320}, it becomes possible to control the coupling strength $g(t)$, thereby \textit{regulating the dissipation rate} $\gamma_{a(b)}(t)$ at each connection point. When the dissipation of the two GAs satisfy the following function~\cite{PhysRevA.84.042341}:
    
    \begin{equation}
        \Gamma_a(t) = \Gamma_b(\tau-t) = 
    \begin{cases}
    	\displaystyle \frac{\Gamma_{\max} e^{\Gamma_{\max} (t-\tau)}}{2 - e^{\Gamma_{\max} (t-\tau)}}, &  t<\tau \\
    	\Gamma_{\max}, & t \geqslant \tau
    \end{cases}
    \end{equation}
    
    the evolution of the system satisfies the dark state condition~\cite{PhysRevLett.118.133601} thereby achieving a lossless unidirectional transfer of energy between the charger and the battery (see the Supplemental Material~\cite{YKX250527} for details). The parameter $\tau$ represents the propagation time and is approximately given by $\tau \approx L_{ab}/v_g$. Figures~\ref{F8}{\color{blue}(b)-(d)} show the characteristics of the battery during the charging process. It can be observed that once the battery is fully charged, its energy no longer fluctuates, indicating a unidirectional transfer of energy between the charger and the battery. Concurrently, the battery’s energy remains isolated from the environment, as the dark state condition is satisfied. \textit{When energy extraction from the battery is required, the phase conditions can be adjusted to reverse the direction of energy flow}, enabling a unidirectional transfer from the battery back to the charger.

       \textit{Conclusion.}---The giant-atom waveguide-QED architecture has been experimentally demonstrated across multiple quantum platforms, exemplified by implementations such as superconducting qubits coupled to surface acoustic waves (SAW)~\cite{Andersson2019,Gustafsson2014,Manenti2017} and ferromagnetic spin ensembles interfaced with a meandering waveguide~\cite{Wang2022}. The QB protocol we proposed can be implemented in superconducting circuits~\cite{Gu2017,Kockum2019}. Two frequency-tunable transmon qubits are coupled to a superconducting resonator for a dispersive readout~\cite{PhysRevA.69.062320}. Both qubits have two connections to a $50$ $\Omega$ coplanar waveguide. The frequency of the qubit are set at $\omega_{0}/2\pi=4$ GHz, the dissipation rate of connection points are $\gamma/2\pi=4$ MHz, and the phase accumulated by the propagation of the electromagnetic field in the waveguide between neighbouring connection points is $\theta=\pi/2$. The maximum power $\mathcal{P}_{\rm max}/2\pi$ and maximum energy $\mathcal{E}_{\rm max}/2\pi$ of decoherence-immune charging between two braided GAs are 0.72 MHz and 4 GHz, respectively.
	
In conclusion, we explore the charging performance of three configurations of GAs when used as QBs. For the braided configuration, the existence of decoherence-immune interactions between the two GAs allows this configuration to achieve \textit{lossless} energy transfer between the charger and the battery, avoiding the \textit{ageing} effect of the battery. The other two configurations, separated and nested, are far inferior to the braided configuration in charging performance due to the absence of such special properties. Furthermore, we have developed a long-range chiral charging protocol utilizing separated GAs. This protocol enables flexible control over the storage and extraction of energy from the battery, while remaining robust to external environmental interference. We hope that our investigation into the properties of GA QBs will contribute to the advancement and realization of future QB technologies.

       \textit{Acknowledgments.}---We express sincere gratitude to Lei Du and Xin Wang for the insightful discussions and valuable suggestions. Y.-H.C. was supported by the National Natural Science Foundation of China under Grant No. 12304390, the Fujian 100 Talents Program, and the Fujian Minjiang Scholar Program. Y. X. was supported by the National Natural Science Foundation of China under Grant No. 62471143, the Key Program of National Natural Science Foundation of Fujian Province under Grant No. 2024J02008, and the project from Fuzhou University under Grant No. JG2020001-2.


\begin{thebibliography}{70}%
		\makeatletter
		\providecommand \@ifxundefined [1]{%
			\@ifx{#1\undefined}
		}%
		\providecommand \@ifnum [1]{%
			\ifnum #1\expandafter \@firstoftwo
			\else \expandafter \@secondoftwo
			\fi
		}%
		\providecommand \@ifx [1]{%
			\ifx #1\expandafter \@firstoftwo
			\else \expandafter \@secondoftwo
			\fi
		}%
		\providecommand \natexlab [1]{#1}%
		\providecommand \enquote  [1]{``#1''}%
		\providecommand \bibnamefont  [1]{#1}%
		\providecommand \bibfnamefont [1]{#1}%
		\providecommand \citenamefont [1]{#1}%
		\providecommand \href@noop [0]{\@secondoftwo}%
		\providecommand \href [0]{\begingroup \@sanitize@url \@href}%
		\providecommand \@href[1]{\@@startlink{#1}\@@href}%
		\providecommand \@@href[1]{\endgroup#1\@@endlink}%
		\providecommand \@sanitize@url [0]{\catcode `\\12\catcode `\$12\catcode
			`\&12\catcode `\#12\catcode `\^12\catcode `\_12\catcode `\%12\relax}%
		\providecommand \@@startlink[1]{}%
		\providecommand \@@endlink[0]{}%
		\providecommand \url  [0]{\begingroup\@sanitize@url \@url }%
		\providecommand \@url [1]{\endgroup\@href {#1}{\urlprefix }}%
		\providecommand \urlprefix  [0]{URL }%
		\providecommand \Eprint [0]{\href }%
		\providecommand \doibase [0]{https://doi.org/}%
		\providecommand \selectlanguage [0]{\@gobble}%
		\providecommand \bibinfo  [0]{\@secondoftwo}%
		\providecommand \bibfield  [0]{\@secondoftwo}%
		\providecommand \translation [1]{[#1]}%
		\providecommand \BibitemOpen [0]{}%
		\providecommand \bibitemStop [0]{}%
		\providecommand \bibitemNoStop [0]{.\EOS\space}%
		\providecommand \EOS [0]{\spacefactor3000\relax}%
		\providecommand \BibitemShut  [1]{\csname bibitem#1\endcsname}%
		\let\auto@bib@innerbib\@empty
		\bibitem [{\citenamefont {Campaioli}\ \emph {et~al.}(2024)\citenamefont
			{Campaioli}, \citenamefont {Gherardini}, \citenamefont {Quach}, \citenamefont
			{Polini},\ and\ \citenamefont {Andolina}}]{RevModPhys.96.031001}%
		\BibitemOpen
		\bibfield  {author} {\bibinfo {author} {\bibfnamefont {F.}~\bibnamefont
				{Campaioli}}, \bibinfo {author} {\bibfnamefont {S.}~\bibnamefont
				{Gherardini}}, \bibinfo {author} {\bibfnamefont {J.~Q.}\ \bibnamefont
				{Quach}}, \bibinfo {author} {\bibfnamefont {M.}~\bibnamefont {Polini}},\ and\
			\bibinfo {author} {\bibfnamefont {G.~M.}\ \bibnamefont {Andolina}},\
		}\bibfield  {title} {\bibinfo {title} {Colloquium: Quantum batteries},\
		}\href {https://doi.org/10.1103/RevModPhys.96.031001} {\bibfield  {journal}
			{\bibinfo  {journal} {Rev. Mod. Phys.}\ }\textbf {\bibinfo {volume} {96}},\
			\bibinfo {pages} {031001} (\bibinfo {year} {2024})}\BibitemShut {NoStop}%
		\bibitem [{\citenamefont {Song}\ \emph {et~al.}(2025)\citenamefont {Song},
			\citenamefont {Wang}, \citenamefont {Zhou}, \citenamefont {Yang},\ and\
			\citenamefont {An}}]{d9k1-75d4}%
		\BibitemOpen
		\bibfield  {author} {\bibinfo {author} {\bibfnamefont {W.-L.}\ \bibnamefont
				{Song}}, \bibinfo {author} {\bibfnamefont {J.-L.}\ \bibnamefont {Wang}},
			\bibinfo {author} {\bibfnamefont {B.}~\bibnamefont {Zhou}}, \bibinfo {author}
			{\bibfnamefont {W.-L.}\ \bibnamefont {Yang}},\ and\ \bibinfo {author}
			{\bibfnamefont {J.-H.}\ \bibnamefont {An}},\ }\bibfield  {title} {\bibinfo
			{title} {Self-discharging mitigated quantum battery},\ }\href
		{https://doi.org/10.1103/d9k1-75d4} {\bibfield  {journal} {\bibinfo
				{journal} {Phys. Rev. Lett.}\ }\textbf {\bibinfo {volume} {135}},\ \bibinfo
			{pages} {020405} (\bibinfo {year} {2025})}\BibitemShut {NoStop}%
		\bibitem [{\citenamefont {Barra}(2019)}]{PhysRevLett.122.210601}%
		\BibitemOpen
		\bibfield  {author} {\bibinfo {author} {\bibfnamefont {F.}~\bibnamefont
				{Barra}},\ }\bibfield  {title} {\bibinfo {title} {Dissipative charging of a
				quantum battery},\ }\href {https://doi.org/10.1103/PhysRevLett.122.210601}
		{\bibfield  {journal} {\bibinfo  {journal} {Phys. Rev. Lett.}\ }\textbf
			{\bibinfo {volume} {122}},\ \bibinfo {pages} {210601} (\bibinfo {year}
			{2019})}\BibitemShut {NoStop}%
		\bibitem [{\citenamefont {Catalano}\ \emph {et~al.}(2024)\citenamefont
			{Catalano}, \citenamefont {Giampaolo}, \citenamefont {Morsch}, \citenamefont
			{Giovannetti},\ and\ \citenamefont {Franchini}}]{PRXQuantum.5.030319}%
		\BibitemOpen
		\bibfield  {author} {\bibinfo {author} {\bibfnamefont {A.}~\bibnamefont
				{Catalano}}, \bibinfo {author} {\bibfnamefont {S.}~\bibnamefont {Giampaolo}},
			\bibinfo {author} {\bibfnamefont {O.}~\bibnamefont {Morsch}}, \bibinfo
			{author} {\bibfnamefont {V.}~\bibnamefont {Giovannetti}},\ and\ \bibinfo
			{author} {\bibfnamefont {F.}~\bibnamefont {Franchini}},\ }\bibfield  {title}
		{\bibinfo {title} {Frustrating quantum batteries},\ }\href
		{https://doi.org/10.1103/PRXQuantum.5.030319} {\bibfield  {journal} {\bibinfo
				{journal} {PRX Quantum}\ }\textbf {\bibinfo {volume} {5}},\ \bibinfo {pages}
			{030319} (\bibinfo {year} {2024})}\BibitemShut {NoStop}%
		\bibitem [{\citenamefont {Ahmadi}\ \emph {et~al.}(2024)\citenamefont {Ahmadi},
			\citenamefont {Mazurek}, \citenamefont {Horodecki},\ and\ \citenamefont
			{Barzanjeh}}]{PhysRevLett.132.210402}%
		\BibitemOpen
		\bibfield  {author} {\bibinfo {author} {\bibfnamefont {B.}~\bibnamefont
				{Ahmadi}}, \bibinfo {author} {\bibfnamefont {P.}~\bibnamefont {Mazurek}},
			\bibinfo {author} {\bibfnamefont {P.}~\bibnamefont {Horodecki}},\ and\
			\bibinfo {author} {\bibfnamefont {S.}~\bibnamefont {Barzanjeh}},\ }\bibfield
		{title} {\bibinfo {title} {Nonreciprocal quantum batteries},\ }\href
		{https://doi.org/10.1103/PhysRevLett.132.210402} {\bibfield  {journal}
			{\bibinfo  {journal} {Phys. Rev. Lett.}\ }\textbf {\bibinfo {volume} {132}},\
			\bibinfo {pages} {210402} (\bibinfo {year} {2024})}\BibitemShut {NoStop}%
		\bibitem [{\citenamefont {Lu}\ \emph {et~al.}(2025)\citenamefont {Lu},
			\citenamefont {Tian}, \citenamefont {L\"u},\ and\ \citenamefont
			{Shang}}]{PhysRevLett.134.180401}%
		\BibitemOpen
		\bibfield  {author} {\bibinfo {author} {\bibfnamefont {Z.-G.}\ \bibnamefont
				{Lu}}, \bibinfo {author} {\bibfnamefont {G.}~\bibnamefont {Tian}}, \bibinfo
			{author} {\bibfnamefont {X.-Y.}\ \bibnamefont {L\"u}},\ and\ \bibinfo
			{author} {\bibfnamefont {C.}~\bibnamefont {Shang}},\ }\bibfield  {title}
		{\bibinfo {title} {Topological quantum batteries},\ }\href
		{https://doi.org/10.1103/PhysRevLett.134.180401} {\bibfield  {journal}
			{\bibinfo  {journal} {Phys. Rev. Lett.}\ }\textbf {\bibinfo {volume} {134}},\
			\bibinfo {pages} {180401} (\bibinfo {year} {2025})}\BibitemShut {NoStop}%
		\bibitem [{\citenamefont {Caravelli}\ \emph {et~al.}(2020)\citenamefont
			{Caravelli}, \citenamefont {Coulter-De~Wit}, \citenamefont
			{Garc\'{\i}a-Pintos},\ and\ \citenamefont
			{Hamma}}]{PhysRevResearch.2.023095}%
		\BibitemOpen
		\bibfield  {author} {\bibinfo {author} {\bibfnamefont {F.}~\bibnamefont
				{Caravelli}}, \bibinfo {author} {\bibfnamefont {G.}~\bibnamefont
				{Coulter-De~Wit}}, \bibinfo {author} {\bibfnamefont {L.~P.}\ \bibnamefont
				{Garc\'{\i}a-Pintos}},\ and\ \bibinfo {author} {\bibfnamefont
				{A.}~\bibnamefont {Hamma}},\ }\bibfield  {title} {\bibinfo {title} {Random
				quantum batteries},\ }\href
		{https://doi.org/10.1103/PhysRevResearch.2.023095} {\bibfield  {journal}
			{\bibinfo  {journal} {Phys. Rev. Res.}\ }\textbf {\bibinfo {volume} {2}},\
			\bibinfo {pages} {023095} (\bibinfo {year} {2020})}\BibitemShut {NoStop}%
		\bibitem [{\citenamefont {Hovhannisyan}\ \emph {et~al.}(2013)\citenamefont
			{Hovhannisyan}, \citenamefont {Perarnau-Llobet}, \citenamefont {Huber},\ and\
			\citenamefont {Ac\'{\i}n}}]{PhysRevLett.111.240401}%
		\BibitemOpen
		\bibfield  {author} {\bibinfo {author} {\bibfnamefont {K.~V.}\ \bibnamefont
				{Hovhannisyan}}, \bibinfo {author} {\bibfnamefont {M.}~\bibnamefont
				{Perarnau-Llobet}}, \bibinfo {author} {\bibfnamefont {M.}~\bibnamefont
				{Huber}},\ and\ \bibinfo {author} {\bibfnamefont {A.}~\bibnamefont
				{Ac\'{\i}n}},\ }\bibfield  {title} {\bibinfo {title} {Entanglement generation
				is not necessary for optimal work extraction},\ }\href
		{https://doi.org/10.1103/PhysRevLett.111.240401} {\bibfield  {journal}
			{\bibinfo  {journal} {Phys. Rev. Lett.}\ }\textbf {\bibinfo {volume} {111}},\
			\bibinfo {pages} {240401} (\bibinfo {year} {2013})}\BibitemShut {NoStop}%
		\bibitem [{\citenamefont {Binder}\ \emph {et~al.}(2015)\citenamefont {Binder},
			\citenamefont {Vinjanampathy}, \citenamefont {Modi},\ and\ \citenamefont
			{Goold}}]{Binder2015}%
		\BibitemOpen
		\bibfield  {author} {\bibinfo {author} {\bibfnamefont {F.~C.}\ \bibnamefont
				{Binder}}, \bibinfo {author} {\bibfnamefont {S.}~\bibnamefont
				{Vinjanampathy}}, \bibinfo {author} {\bibfnamefont {K.}~\bibnamefont
				{Modi}},\ and\ \bibinfo {author} {\bibfnamefont {J.}~\bibnamefont {Goold}},\
		}\bibfield  {title} {\bibinfo {title} {Quantacell: powerful charging of
				quantum batteries},\ }\href {https://doi.org/10.1088/1367-2630/17/7/075015}
		{\bibfield  {journal} {\bibinfo  {journal} {New J. Phys.}\ }\textbf {\bibinfo
				{volume} {17}},\ \bibinfo {pages} {075015} (\bibinfo {year}
			{2015})}\BibitemShut {NoStop}%
		\bibitem [{\citenamefont {Andolina}\ \emph {et~al.}(2019)\citenamefont
			{Andolina}, \citenamefont {Keck}, \citenamefont {Mari}, \citenamefont
			{Campisi}, \citenamefont {Giovannetti},\ and\ \citenamefont
			{Polini}}]{PhysRevLett.122.047702}%
		\BibitemOpen
		\bibfield  {author} {\bibinfo {author} {\bibfnamefont {G.~M.}\ \bibnamefont
				{Andolina}}, \bibinfo {author} {\bibfnamefont {M.}~\bibnamefont {Keck}},
			\bibinfo {author} {\bibfnamefont {A.}~\bibnamefont {Mari}}, \bibinfo {author}
			{\bibfnamefont {M.}~\bibnamefont {Campisi}}, \bibinfo {author} {\bibfnamefont
				{V.}~\bibnamefont {Giovannetti}},\ and\ \bibinfo {author} {\bibfnamefont
				{M.}~\bibnamefont {Polini}},\ }\bibfield  {title} {\bibinfo {title}
			{Extractable work, the role of correlations, and asymptotic freedom in
				quantum batteries},\ }\href {https://doi.org/10.1103/PhysRevLett.122.047702}
		{\bibfield  {journal} {\bibinfo  {journal} {Phys. Rev. Lett.}\ }\textbf
			{\bibinfo {volume} {122}},\ \bibinfo {pages} {047702} (\bibinfo {year}
			{2019})}\BibitemShut {NoStop}%
		\bibitem [{\citenamefont {Ferraro}\ \emph {et~al.}(2018)\citenamefont
			{Ferraro}, \citenamefont {Campisi}, \citenamefont {Andolina}, \citenamefont
			{Pellegrini},\ and\ \citenamefont {Polini}}]{PhysRevLett.120.117702}%
		\BibitemOpen
		\bibfield  {author} {\bibinfo {author} {\bibfnamefont {D.}~\bibnamefont
				{Ferraro}}, \bibinfo {author} {\bibfnamefont {M.}~\bibnamefont {Campisi}},
			\bibinfo {author} {\bibfnamefont {G.~M.}\ \bibnamefont {Andolina}}, \bibinfo
			{author} {\bibfnamefont {V.}~\bibnamefont {Pellegrini}},\ and\ \bibinfo
			{author} {\bibfnamefont {M.}~\bibnamefont {Polini}},\ }\bibfield  {title}
		{\bibinfo {title} {High-power collective charging of a solid-state quantum
				battery},\ }\href {https://doi.org/10.1103/PhysRevLett.120.117702} {\bibfield
			{journal} {\bibinfo  {journal} {Phys. Rev. Lett.}\ }\textbf {\bibinfo
				{volume} {120}},\ \bibinfo {pages} {117702} (\bibinfo {year}
			{2018})}\BibitemShut {NoStop}%
		\bibitem [{\citenamefont {Crescente}\ \emph {et~al.}(2020)\citenamefont
			{Crescente}, \citenamefont {Carrega}, \citenamefont {Sassetti},\ and\
			\citenamefont {Ferraro}}]{PhysRevB.102.245407}%
		\BibitemOpen
		\bibfield  {author} {\bibinfo {author} {\bibfnamefont {A.}~\bibnamefont
				{Crescente}}, \bibinfo {author} {\bibfnamefont {M.}~\bibnamefont {Carrega}},
			\bibinfo {author} {\bibfnamefont {M.}~\bibnamefont {Sassetti}},\ and\
			\bibinfo {author} {\bibfnamefont {D.}~\bibnamefont {Ferraro}},\ }\bibfield
		{title} {\bibinfo {title} {Ultrafast charging in a two-photon {Dicke} quantum
				battery},\ }\href {https://doi.org/10.1103/PhysRevB.102.245407} {\bibfield
			{journal} {\bibinfo  {journal} {Phys. Rev. B}\ }\textbf {\bibinfo {volume}
				{102}},\ \bibinfo {pages} {245407} (\bibinfo {year} {2020})}\BibitemShut
		{NoStop}%
		\bibitem [{\citenamefont {Chen}\ \emph {et~al.}(2020)\citenamefont {Chen},
			\citenamefont {Zhan}, \citenamefont {Shao}, \citenamefont {Zhang},
			\citenamefont {Zhang},\ and\ \citenamefont {Wang}}]{Chen2020}%
		\BibitemOpen
		\bibfield  {author} {\bibinfo {author} {\bibfnamefont {J.}~\bibnamefont
				{Chen}}, \bibinfo {author} {\bibfnamefont {L.}~\bibnamefont {Zhan}}, \bibinfo
			{author} {\bibfnamefont {L.}~\bibnamefont {Shao}}, \bibinfo {author}
			{\bibfnamefont {X.}~\bibnamefont {Zhang}}, \bibinfo {author} {\bibfnamefont
				{Y.}~\bibnamefont {Zhang}},\ and\ \bibinfo {author} {\bibfnamefont
				{X.}~\bibnamefont {Wang}},\ }\bibfield  {title} {\bibinfo {title} {Charging
				quantum batteries with a general harmonic driving field},\ }\href
		{http://dx.doi.org/10.1002/andp.201900487} {\bibfield  {journal} {\bibinfo
				{journal} {Ann. Phys.}\ }\textbf {\bibinfo {volume} {532}} (\bibinfo {year}
			{2020})}\BibitemShut {NoStop}%
		\bibitem [{\citenamefont {Peng}\ \emph {et~al.}(2021)\citenamefont {Peng},
			\citenamefont {He}, \citenamefont {Chesi}, \citenamefont {Lin},\ and\
			\citenamefont {Guan}}]{PhysRevA.103.052220}%
		\BibitemOpen
		\bibfield  {author} {\bibinfo {author} {\bibfnamefont {L.}~\bibnamefont
				{Peng}}, \bibinfo {author} {\bibfnamefont {W.-B.}\ \bibnamefont {He}},
			\bibinfo {author} {\bibfnamefont {S.}~\bibnamefont {Chesi}}, \bibinfo
			{author} {\bibfnamefont {H.-Q.}\ \bibnamefont {Lin}},\ and\ \bibinfo {author}
			{\bibfnamefont {X.-W.}\ \bibnamefont {Guan}},\ }\bibfield  {title} {\bibinfo
			{title} {Lower and upper bounds of quantum battery power in multiple central
				spin systems},\ }\href {https://doi.org/10.1103/PhysRevA.103.052220}
		{\bibfield  {journal} {\bibinfo  {journal} {Phys. Rev. A}\ }\textbf {\bibinfo
				{volume} {103}},\ \bibinfo {pages} {052220} (\bibinfo {year}
			{2021})}\BibitemShut {NoStop}%
		\bibitem [{\citenamefont {Juli\`a-Farr\'e}\ \emph {et~al.}(2020)\citenamefont
			{Juli\`a-Farr\'e}, \citenamefont {Salamon}, \citenamefont {Riera},
			\citenamefont {Bera},\ and\ \citenamefont
			{Lewenstein}}]{PhysRevResearch.2.023113}%
		\BibitemOpen
		\bibfield  {author} {\bibinfo {author} {\bibfnamefont {S.}~\bibnamefont
				{Juli\`a-Farr\'e}}, \bibinfo {author} {\bibfnamefont {T.}~\bibnamefont
				{Salamon}}, \bibinfo {author} {\bibfnamefont {A.}~\bibnamefont {Riera}},
			\bibinfo {author} {\bibfnamefont {M.~N.}\ \bibnamefont {Bera}},\ and\
			\bibinfo {author} {\bibfnamefont {M.}~\bibnamefont {Lewenstein}},\ }\bibfield
		{title} {\bibinfo {title} {Bounds on the capacity and power of quantum
				batteries},\ }\href {https://doi.org/10.1103/PhysRevResearch.2.023113}
		{\bibfield  {journal} {\bibinfo  {journal} {Phys. Rev. Res.}\ }\textbf
			{\bibinfo {volume} {2}},\ \bibinfo {pages} {023113} (\bibinfo {year}
			{2020})}\BibitemShut {NoStop}%
		\bibitem [{\citenamefont {Dias}\ \emph {et~al.}(2024)\citenamefont {Dias},
			\citenamefont {Wang}, \citenamefont {Nemoto}, \citenamefont {Franco},\ and\
			\citenamefont {Munro}}]{Jose2024}%
		\BibitemOpen
		\bibfield  {author} {\bibinfo {author} {\bibfnamefont {J.}~\bibnamefont
				{Dias}}, \bibinfo {author} {\bibfnamefont {H.}~\bibnamefont {Wang}}, \bibinfo
			{author} {\bibfnamefont {K.}~\bibnamefont {Nemoto}}, \bibinfo {author}
			{\bibfnamefont {N.}~\bibnamefont {Franco}},\ and\ \bibinfo {author}
			{\bibfnamefont {W.~J.}\ \bibnamefont {Munro}},\ }\bibfield  {title} {\bibinfo
			{title} {Efficient charging of multiple open quantum batteries through
				dissipation and pumping},\ }\href {https://arxiv.org/pdf/2410.19303}
		{\bibfield  {journal} {\bibinfo  {journal} {arXiv:2410.19303}\ } (\bibinfo
			{year} {2024})}\BibitemShut {NoStop}%
		\bibitem [{\citenamefont {Campaioli}\ \emph {et~al.}(2017)\citenamefont
			{Campaioli}, \citenamefont {Pollock}, \citenamefont {Binder}, \citenamefont
			{C\'eleri}, \citenamefont {Goold}, \citenamefont {Vinjanampathy},\ and\
			\citenamefont {Modi}}]{PhysRevLett.118.150601}%
		\BibitemOpen
		\bibfield  {author} {\bibinfo {author} {\bibfnamefont {F.}~\bibnamefont
				{Campaioli}}, \bibinfo {author} {\bibfnamefont {F.~A.}\ \bibnamefont
				{Pollock}}, \bibinfo {author} {\bibfnamefont {F.~C.}\ \bibnamefont {Binder}},
			\bibinfo {author} {\bibfnamefont {L.}~\bibnamefont {C\'eleri}}, \bibinfo
			{author} {\bibfnamefont {J.}~\bibnamefont {Goold}}, \bibinfo {author}
			{\bibfnamefont {S.}~\bibnamefont {Vinjanampathy}},\ and\ \bibinfo {author}
			{\bibfnamefont {K.}~\bibnamefont {Modi}},\ }\bibfield  {title} {\bibinfo
			{title} {Enhancing the charging power of quantum batteries},\ }\href
		{https://doi.org/10.1103/PhysRevLett.118.150601} {\bibfield  {journal}
			{\bibinfo  {journal} {Phys. Rev. Lett.}\ }\textbf {\bibinfo {volume} {118}},\
			\bibinfo {pages} {150601} (\bibinfo {year} {2017})}\BibitemShut {NoStop}%
		\bibitem [{\citenamefont {Wang}\ \emph
			{et~al.}(2025{\natexlab{a}})\citenamefont {Wang}, \citenamefont {Huang},\
			and\ \citenamefont {Zhang}}]{WangYI2025}%
		\BibitemOpen
		\bibfield  {author} {\bibinfo {author} {\bibfnamefont {Y.}~\bibnamefont
				{Wang}}, \bibinfo {author} {\bibfnamefont {X.}~\bibnamefont {Huang}},\ and\
			\bibinfo {author} {\bibfnamefont {T.}~\bibnamefont {Zhang}},\ }\bibfield
		{title} {\bibinfo {title} {Distribution relationship of quantum battery
				capacity},\ }\href {http://dx.doi.org/10.1002/qute.202400652} {\bibfield
			{journal} {\bibinfo  {journal} {Adv. Quantum Technol.}\ } (\bibinfo {year}
			{2025}{\natexlab{a}})}\BibitemShut {NoStop}%
		\bibitem [{\citenamefont {Wang}\ \emph
			{et~al.}(2025{\natexlab{b}})\citenamefont {Wang}, \citenamefont {Ge},
			\citenamefont {Zhang}, \citenamefont {Fei}, \citenamefont {Gao},\ and\
			\citenamefont {Wang}}]{WangYA2025}%
		\BibitemOpen
		\bibfield  {author} {\bibinfo {author} {\bibfnamefont {Y.-K.}\ \bibnamefont
				{Wang}}, \bibinfo {author} {\bibfnamefont {L.-Z.}\ \bibnamefont {Ge}},
			\bibinfo {author} {\bibfnamefont {T.}~\bibnamefont {Zhang}}, \bibinfo
			{author} {\bibfnamefont {S.-M.}\ \bibnamefont {Fei}}, \bibinfo {author}
			{\bibfnamefont {Y.}~\bibnamefont {Gao}},\ and\ \bibinfo {author}
			{\bibfnamefont {Z.-X.}\ \bibnamefont {Wang}},\ }\bibfield  {title} {\bibinfo
			{title} {Dynamics of quantum battery capacity under {Markovian} channels},\
		}\href {http://dx.doi.org/10.1007/s11128-025-04645-5} {\bibfield  {journal}
			{\bibinfo  {journal} {Quantum Inf. Process.}\ }\textbf {\bibinfo {volume}
				{24}} (\bibinfo {year} {2025}{\natexlab{b}})}\BibitemShut {NoStop}%
		\bibitem [{\citenamefont {Yang}\ \emph {et~al.}(2023)\citenamefont {Yang},
			\citenamefont {Yang}, \citenamefont {Alimuddin}, \citenamefont {Salvia},
			\citenamefont {Fei}, \citenamefont {Zhao}, \citenamefont {Nimmrichter},\ and\
			\citenamefont {Luo}}]{PhysRevLett.131.030402}%
		\BibitemOpen
		\bibfield  {author} {\bibinfo {author} {\bibfnamefont {X.}~\bibnamefont
				{Yang}}, \bibinfo {author} {\bibfnamefont {Y.-H.}\ \bibnamefont {Yang}},
			\bibinfo {author} {\bibfnamefont {M.}~\bibnamefont {Alimuddin}}, \bibinfo
			{author} {\bibfnamefont {R.}~\bibnamefont {Salvia}}, \bibinfo {author}
			{\bibfnamefont {S.-M.}\ \bibnamefont {Fei}}, \bibinfo {author} {\bibfnamefont
				{L.-M.}\ \bibnamefont {Zhao}}, \bibinfo {author} {\bibfnamefont
				{S.}~\bibnamefont {Nimmrichter}},\ and\ \bibinfo {author} {\bibfnamefont
				{M.-X.}\ \bibnamefont {Luo}},\ }\bibfield  {title} {\bibinfo {title} {Battery
				capacity of energy-storing quantum systems},\ }\href
		{https://doi.org/10.1103/PhysRevLett.131.030402} {\bibfield  {journal}
			{\bibinfo  {journal} {Phys. Rev. Lett.}\ }\textbf {\bibinfo {volume} {131}},\
			\bibinfo {pages} {030402} (\bibinfo {year} {2023})}\BibitemShut {NoStop}%
		\bibitem [{\citenamefont {Francica}\ \emph {et~al.}(2020)\citenamefont
			{Francica}, \citenamefont {Binder}, \citenamefont {Guarnieri}, \citenamefont
			{Mitchison}, \citenamefont {Goold},\ and\ \citenamefont
			{Plastina}}]{PhysRevLett.125.180603}%
		\BibitemOpen
		\bibfield  {author} {\bibinfo {author} {\bibfnamefont {G.}~\bibnamefont
				{Francica}}, \bibinfo {author} {\bibfnamefont {F.~C.}\ \bibnamefont
				{Binder}}, \bibinfo {author} {\bibfnamefont {G.}~\bibnamefont {Guarnieri}},
			\bibinfo {author} {\bibfnamefont {M.~T.}\ \bibnamefont {Mitchison}}, \bibinfo
			{author} {\bibfnamefont {J.}~\bibnamefont {Goold}},\ and\ \bibinfo {author}
			{\bibfnamefont {F.}~\bibnamefont {Plastina}},\ }\bibfield  {title} {\bibinfo
			{title} {Quantum coherence and ergotropy},\ }\href
		{https://doi.org/10.1103/PhysRevLett.125.180603} {\bibfield  {journal}
			{\bibinfo  {journal} {Phys. Rev. Lett.}\ }\textbf {\bibinfo {volume} {125}},\
			\bibinfo {pages} {180603} (\bibinfo {year} {2020})}\BibitemShut {NoStop}%
		\bibitem [{\citenamefont {Shi}\ \emph {et~al.}(2022)\citenamefont {Shi},
			\citenamefont {Ding}, \citenamefont {Wan}, \citenamefont {Wang},\ and\
			\citenamefont {Yang}}]{PhysRevLett.129.130602}%
		\BibitemOpen
		\bibfield  {author} {\bibinfo {author} {\bibfnamefont {H.-L.}\ \bibnamefont
				{Shi}}, \bibinfo {author} {\bibfnamefont {S.}~\bibnamefont {Ding}}, \bibinfo
			{author} {\bibfnamefont {Q.-K.}\ \bibnamefont {Wan}}, \bibinfo {author}
			{\bibfnamefont {X.-H.}\ \bibnamefont {Wang}},\ and\ \bibinfo {author}
			{\bibfnamefont {W.-L.}\ \bibnamefont {Yang}},\ }\bibfield  {title} {\bibinfo
			{title} {Entanglement, coherence, and extractable work in quantum
				batteries},\ }\href {https://doi.org/10.1103/PhysRevLett.129.130602}
		{\bibfield  {journal} {\bibinfo  {journal} {Phys. Rev. Lett.}\ }\textbf
			{\bibinfo {volume} {129}},\ \bibinfo {pages} {130602} (\bibinfo {year}
			{2022})}\BibitemShut {NoStop}%
		\bibitem [{\citenamefont {Çakmak}(2020)}]{PhysRevE.102.042111}%
		\BibitemOpen
		\bibfield  {author} {\bibinfo {author} {\bibfnamefont {B.}~\bibnamefont
				{Çakmak}},\ }\bibfield  {title} {\bibinfo {title} {Ergotropy from coherences
				in an open quantum system},\ }\href
		{https://doi.org/10.1103/PhysRevE.102.042111} {\bibfield  {journal} {\bibinfo
				{journal} {Phys. Rev. E}\ }\textbf {\bibinfo {volume} {102}},\ \bibinfo
			{pages} {042111} (\bibinfo {year} {2020})}\BibitemShut {NoStop}%
		\bibitem [{\citenamefont {Pirmoradian}\ and\ \citenamefont
			{M\o{}lmer}(2019)}]{PhysRevA.100.043833}%
		\BibitemOpen
		\bibfield  {author} {\bibinfo {author} {\bibfnamefont {F.}~\bibnamefont
				{Pirmoradian}}\ and\ \bibinfo {author} {\bibfnamefont {K.}~\bibnamefont
				{M\o{}lmer}},\ }\bibfield  {title} {\bibinfo {title} {Aging of a quantum
				battery},\ }\href {https://doi.org/10.1103/PhysRevA.100.043833} {\bibfield
			{journal} {\bibinfo  {journal} {Phys. Rev. A}\ }\textbf {\bibinfo {volume}
				{100}},\ \bibinfo {pages} {043833} (\bibinfo {year} {2019})}\BibitemShut
		{NoStop}%
		\bibitem [{\citenamefont {Liu}\ \emph {et~al.}(2019)\citenamefont {Liu},
			\citenamefont {Segal},\ and\ \citenamefont {Hanna}}]{LSH2019}%
		\BibitemOpen
		\bibfield  {author} {\bibinfo {author} {\bibfnamefont {J.}~\bibnamefont
				{Liu}}, \bibinfo {author} {\bibfnamefont {D.}~\bibnamefont {Segal}},\ and\
			\bibinfo {author} {\bibfnamefont {G.}~\bibnamefont {Hanna}},\ }\bibfield
		{title} {\bibinfo {title} {Loss-free excitonic quantum battery},\ }\href
		{https://doi.org/10.1021/acs.jpcc.9b06373} {\bibfield  {journal} {\bibinfo
				{journal} {J. Phys. Chem. C}\ }\textbf {\bibinfo {volume} {123}},\ \bibinfo
			{pages} {18303} (\bibinfo {year} {2019})}\BibitemShut {NoStop}%
		\bibitem [{\citenamefont {Quach}\ and\ \citenamefont
			{Munro}(2020)}]{PhysRevApplied.14.024092}%
		\BibitemOpen
		\bibfield  {author} {\bibinfo {author} {\bibfnamefont {J.~Q.}\ \bibnamefont
				{Quach}}\ and\ \bibinfo {author} {\bibfnamefont {W.~J.}\ \bibnamefont
				{Munro}},\ }\bibfield  {title} {\bibinfo {title} {Using dark states to charge
				and stabilize open quantum batteries},\ }\href
		{https://doi.org/10.1103/PhysRevApplied.14.024092} {\bibfield  {journal}
			{\bibinfo  {journal} {Phys. Rev. Appl.}\ }\textbf {\bibinfo {volume} {14}},\
			\bibinfo {pages} {024092} (\bibinfo {year} {2020})}\BibitemShut {NoStop}%
		\bibitem [{\citenamefont {Bai}\ and\ \citenamefont
			{An}(2020)}]{PhysRevA.102.060201}%
		\BibitemOpen
		\bibfield  {author} {\bibinfo {author} {\bibfnamefont {S.-Y.}\ \bibnamefont
				{Bai}}\ and\ \bibinfo {author} {\bibfnamefont {J.-H.}\ \bibnamefont {An}},\
		}\bibfield  {title} {\bibinfo {title} {Floquet engineering to reactivate a
				dissipative quantum battery},\ }\href
		{https://doi.org/10.1103/PhysRevA.102.060201} {\bibfield  {journal} {\bibinfo
				{journal} {Phys. Rev. A}\ }\textbf {\bibinfo {volume} {102}},\ \bibinfo
			{pages} {060201} (\bibinfo {year} {2020})}\BibitemShut {NoStop}%
		\bibitem [{\citenamefont {Mitchison}\ \emph {et~al.}(2021)\citenamefont
			{Mitchison}, \citenamefont {Goold},\ and\ \citenamefont
			{Prior}}]{Mitchison2021}%
		\BibitemOpen
		\bibfield  {author} {\bibinfo {author} {\bibfnamefont {M.~T.}\ \bibnamefont
				{Mitchison}}, \bibinfo {author} {\bibfnamefont {J.}~\bibnamefont {Goold}},\
			and\ \bibinfo {author} {\bibfnamefont {J.}~\bibnamefont {Prior}},\ }\bibfield
		{title} {\bibinfo {title} {Charging a quantum battery with linear feedback
				control},\ }\href {https://doi.org/10.22331/q-2021-07-13-500} {\bibfield
			{journal} {\bibinfo  {journal} {Quantum}\ }\textbf {\bibinfo {volume} {5}},\
			\bibinfo {pages} {500} (\bibinfo {year} {2021})}\BibitemShut {NoStop}%
		\bibitem [{\citenamefont {Xu}\ \emph {et~al.}(2022)\citenamefont {Xu},
			\citenamefont {Li}, \citenamefont {Li}, \citenamefont {Zhu}, \citenamefont
			{Zhang},\ and\ \citenamefont {Liu}}]{PhysRevA.106.012425}%
		\BibitemOpen
		\bibfield  {author} {\bibinfo {author} {\bibfnamefont {K.}~\bibnamefont
				{Xu}}, \bibinfo {author} {\bibfnamefont {H.-G.}\ \bibnamefont {Li}}, \bibinfo
			{author} {\bibfnamefont {Z.-G.}\ \bibnamefont {Li}}, \bibinfo {author}
			{\bibfnamefont {H.-J.}\ \bibnamefont {Zhu}}, \bibinfo {author} {\bibfnamefont
				{G.-F.}\ \bibnamefont {Zhang}},\ and\ \bibinfo {author} {\bibfnamefont
				{W.-M.}\ \bibnamefont {Liu}},\ }\bibfield  {title} {\bibinfo {title}
			{Charging performance of quantum batteries in a double-layer environment},\
		}\href {https://doi.org/10.1103/PhysRevA.106.012425} {\bibfield  {journal}
			{\bibinfo  {journal} {Phys. Rev. A}\ }\textbf {\bibinfo {volume} {106}},\
			\bibinfo {pages} {012425} (\bibinfo {year} {2022})}\BibitemShut {NoStop}%
		\bibitem [{\citenamefont {Carrasco}\ \emph {et~al.}(2022)\citenamefont
			{Carrasco}, \citenamefont {Maze}, \citenamefont {Hermann-Avigliano},\ and\
			\citenamefont {Barra}}]{PhysRevE.105.064119}%
		\BibitemOpen
		\bibfield  {author} {\bibinfo {author} {\bibfnamefont {J.}~\bibnamefont
				{Carrasco}}, \bibinfo {author} {\bibfnamefont {J.~R.}\ \bibnamefont {Maze}},
			\bibinfo {author} {\bibfnamefont {C.}~\bibnamefont {Hermann-Avigliano}},\
			and\ \bibinfo {author} {\bibfnamefont {F.}~\bibnamefont {Barra}},\ }\bibfield
		{title} {\bibinfo {title} {Collective enhancement in dissipative quantum
				batteries},\ }\href {https://doi.org/10.1103/PhysRevE.105.064119} {\bibfield
			{journal} {\bibinfo  {journal} {Phys. Rev. E}\ }\textbf {\bibinfo {volume}
				{105}},\ \bibinfo {pages} {064119} (\bibinfo {year} {2022})}\BibitemShut
		{NoStop}%
		\bibitem [{\citenamefont {Xu}\ \emph {et~al.}(2021)\citenamefont {Xu},
			\citenamefont {Zhu}, \citenamefont {Zhang},\ and\ \citenamefont
			{Liu}}]{PhysRevE.104.064143}%
		\BibitemOpen
		\bibfield  {author} {\bibinfo {author} {\bibfnamefont {K.}~\bibnamefont
				{Xu}}, \bibinfo {author} {\bibfnamefont {H.-J.}\ \bibnamefont {Zhu}},
			\bibinfo {author} {\bibfnamefont {G.-F.}\ \bibnamefont {Zhang}},\ and\
			\bibinfo {author} {\bibfnamefont {W.-M.}\ \bibnamefont {Liu}},\ }\bibfield
		{title} {\bibinfo {title} {Enhancing the performance of an open quantum
				battery via environment engineering},\ }\href
		{https://doi.org/10.1103/PhysRevE.104.064143} {\bibfield  {journal} {\bibinfo
				{journal} {Phys. Rev. E}\ }\textbf {\bibinfo {volume} {104}},\ \bibinfo
			{pages} {064143} (\bibinfo {year} {2021})}\BibitemShut {NoStop}%
		\bibitem [{\citenamefont {Gherardini}\ \emph {et~al.}(2020)\citenamefont
			{Gherardini}, \citenamefont {Campaioli}, \citenamefont {Caruso},\ and\
			\citenamefont {Binder}}]{PhysRevResearch.2.013095}%
		\BibitemOpen
		\bibfield  {author} {\bibinfo {author} {\bibfnamefont {S.}~\bibnamefont
				{Gherardini}}, \bibinfo {author} {\bibfnamefont {F.}~\bibnamefont
				{Campaioli}}, \bibinfo {author} {\bibfnamefont {F.}~\bibnamefont {Caruso}},\
			and\ \bibinfo {author} {\bibfnamefont {F.~C.}\ \bibnamefont {Binder}},\
		}\bibfield  {title} {\bibinfo {title} {Stabilizing open quantum batteries by
				sequential measurements},\ }\href
		{https://doi.org/10.1103/PhysRevResearch.2.013095} {\bibfield  {journal}
			{\bibinfo  {journal} {Phys. Rev. Res.}\ }\textbf {\bibinfo {volume} {2}},\
			\bibinfo {pages} {013095} (\bibinfo {year} {2020})}\BibitemShut {NoStop}%
		\bibitem [{\citenamefont {Frisk~Kockum}\ \emph {et~al.}(2014)\citenamefont
			{Frisk~Kockum}, \citenamefont {Delsing},\ and\ \citenamefont
			{Johansson}}]{PhysRevA.90.013837}%
		\BibitemOpen
		\bibfield  {author} {\bibinfo {author} {\bibfnamefont {A.}~\bibnamefont
				{Frisk~Kockum}}, \bibinfo {author} {\bibfnamefont {P.}~\bibnamefont
				{Delsing}},\ and\ \bibinfo {author} {\bibfnamefont {G.}~\bibnamefont
				{Johansson}},\ }\bibfield  {title} {\bibinfo {title} {Designing
				frequency-dependent relaxation rates and {Lamb} shifts for a giant artificial
				atom},\ }\href {https://doi.org/10.1103/PhysRevA.90.013837} {\bibfield
			{journal} {\bibinfo  {journal} {Phys. Rev. A}\ }\textbf {\bibinfo {volume}
				{90}},\ \bibinfo {pages} {013837} (\bibinfo {year} {2014})}\BibitemShut
		{NoStop}%
		\bibitem [{\citenamefont {Yan}\ and\ \citenamefont
			{L\"u}(2024)}]{PhysRevResearch.6.013301}%
		\BibitemOpen
		\bibfield  {author} {\bibinfo {author} {\bibfnamefont {Y.}~\bibnamefont
				{Yan}}\ and\ \bibinfo {author} {\bibfnamefont {Z.}~\bibnamefont {L\"u}},\
		}\bibfield  {title} {\bibinfo {title} {Controllable spontaneous emission
				spectrum in an artificial giant atom: Dark lines and bound states},\ }\href
		{https://doi.org/10.1103/PhysRevResearch.6.013301} {\bibfield  {journal}
			{\bibinfo  {journal} {Phys. Rev. Res.}\ }\textbf {\bibinfo {volume} {6}},\
			\bibinfo {pages} {013301} (\bibinfo {year} {2024})}\BibitemShut {NoStop}%
		\bibitem [{\citenamefont {Wang}\ \emph {et~al.}(2024)\citenamefont {Wang},
			\citenamefont {Zhu}, \citenamefont {Liu},\ and\ \citenamefont
			{Nori}}]{PhysRevResearch.6.013279}%
		\BibitemOpen
		\bibfield  {author} {\bibinfo {author} {\bibfnamefont {X.}~\bibnamefont
				{Wang}}, \bibinfo {author} {\bibfnamefont {H.-B.}\ \bibnamefont {Zhu}},
			\bibinfo {author} {\bibfnamefont {T.}~\bibnamefont {Liu}},\ and\ \bibinfo
			{author} {\bibfnamefont {F.}~\bibnamefont {Nori}},\ }\bibfield  {title}
		{\bibinfo {title} {Realizing quantum optics in structured environments with
				giant atoms},\ }\href {https://doi.org/10.1103/PhysRevResearch.6.013279}
		{\bibfield  {journal} {\bibinfo  {journal} {Phys. Rev. Res.}\ }\textbf
			{\bibinfo {volume} {6}},\ \bibinfo {pages} {013279} (\bibinfo {year}
			{2024})}\BibitemShut {NoStop}%
		\bibitem [{\citenamefont {Qiu}\ and\ \citenamefont
			{L\"u}(2024)}]{PhysRevResearch.6.033243}%
		\BibitemOpen
		\bibfield  {author} {\bibinfo {author} {\bibfnamefont {Q.-Y.}\ \bibnamefont
				{Qiu}}\ and\ \bibinfo {author} {\bibfnamefont {X.-Y.}\ \bibnamefont {L\"u}},\
		}\bibfield  {title} {\bibinfo {title} {Non-{Markovian} collective emission of
				giant emitters in the zeno regime},\ }\href
		{https://doi.org/10.1103/PhysRevResearch.6.033243} {\bibfield  {journal}
			{\bibinfo  {journal} {Phys. Rev. Res.}\ }\textbf {\bibinfo {volume} {6}},\
			\bibinfo {pages} {033243} (\bibinfo {year} {2024})}\BibitemShut {NoStop}%
		\bibitem [{\citenamefont {Raaholt~Ingelsten}\ \emph {et~al.}(2024)\citenamefont
			{Raaholt~Ingelsten}, \citenamefont {Kockum},\ and\ \citenamefont
			{Soro}}]{PhysRevResearch.6.043222}%
		\BibitemOpen
		\bibfield  {author} {\bibinfo {author} {\bibfnamefont {E.}~\bibnamefont
				{Raaholt~Ingelsten}}, \bibinfo {author} {\bibfnamefont {A.~F.}\ \bibnamefont
				{Kockum}},\ and\ \bibinfo {author} {\bibfnamefont {A.}~\bibnamefont {Soro}},\
		}\bibfield  {title} {\bibinfo {title} {Avoiding decoherence with giant atoms
				in a two-dimensional structured environment},\ }\href
		{https://doi.org/10.1103/PhysRevResearch.6.043222} {\bibfield  {journal}
			{\bibinfo  {journal} {Phys. Rev. Res.}\ }\textbf {\bibinfo {volume} {6}},\
			\bibinfo {pages} {043222} (\bibinfo {year} {2024})}\BibitemShut {NoStop}%
		\bibitem [{\citenamefont {Chen}\ \emph {et~al.}(2023)\citenamefont {Chen},
			\citenamefont {Du}, \citenamefont {Zhang}, \citenamefont {Guo}, \citenamefont
			{Wu}, \citenamefont {Artoni},\ and\ \citenamefont
			{La~Rocca}}]{PhysRevResearch.5.043135}%
		\BibitemOpen
		\bibfield  {author} {\bibinfo {author} {\bibfnamefont {Y.-T.}\ \bibnamefont
				{Chen}}, \bibinfo {author} {\bibfnamefont {L.}~\bibnamefont {Du}}, \bibinfo
			{author} {\bibfnamefont {Y.}~\bibnamefont {Zhang}}, \bibinfo {author}
			{\bibfnamefont {L.}~\bibnamefont {Guo}}, \bibinfo {author} {\bibfnamefont
				{J.-H.}\ \bibnamefont {Wu}}, \bibinfo {author} {\bibfnamefont
				{M.}~\bibnamefont {Artoni}},\ and\ \bibinfo {author} {\bibfnamefont {G.~C.}\
				\bibnamefont {La~Rocca}},\ }\bibfield  {title} {\bibinfo {title} {Giant-atom
				effects on population and entanglement dynamics of {Rydberg} atoms in the
				optical regime},\ }\href {https://doi.org/10.1103/PhysRevResearch.5.043135}
		{\bibfield  {journal} {\bibinfo  {journal} {Phys. Rev. Res.}\ }\textbf
			{\bibinfo {volume} {5}},\ \bibinfo {pages} {043135} (\bibinfo {year}
			{2023})}\BibitemShut {NoStop}%
		\bibitem [{\citenamefont {Du}\ \emph {et~al.}(2022)\citenamefont {Du},
			\citenamefont {Chen}, \citenamefont {Zhang},\ and\ \citenamefont
			{Li}}]{PhysRevResearch.4.023198}%
		\BibitemOpen
		\bibfield  {author} {\bibinfo {author} {\bibfnamefont {L.}~\bibnamefont
				{Du}}, \bibinfo {author} {\bibfnamefont {Y.-T.}\ \bibnamefont {Chen}},
			\bibinfo {author} {\bibfnamefont {Y.}~\bibnamefont {Zhang}},\ and\ \bibinfo
			{author} {\bibfnamefont {Y.}~\bibnamefont {Li}},\ }\bibfield  {title}
		{\bibinfo {title} {Giant atoms with time-dependent couplings},\ }\href
		{https://doi.org/10.1103/PhysRevResearch.4.023198} {\bibfield  {journal}
			{\bibinfo  {journal} {Phys. Rev. Res.}\ }\textbf {\bibinfo {volume} {4}},\
			\bibinfo {pages} {023198} (\bibinfo {year} {2022})}\BibitemShut {NoStop}%
		\bibitem [{\citenamefont {Carollo}\ \emph {et~al.}(2020)\citenamefont
			{Carollo}, \citenamefont {Cilluffo},\ and\ \citenamefont
			{Ciccarello}}]{PhysRevResearch.2.043184}%
		\BibitemOpen
		\bibfield  {author} {\bibinfo {author} {\bibfnamefont {A.}~\bibnamefont
				{Carollo}}, \bibinfo {author} {\bibfnamefont {D.}~\bibnamefont {Cilluffo}},\
			and\ \bibinfo {author} {\bibfnamefont {F.}~\bibnamefont {Ciccarello}},\
		}\bibfield  {title} {\bibinfo {title} {Mechanism of decoherence-free coupling
				between giant atoms},\ }\href
		{https://doi.org/10.1103/PhysRevResearch.2.043184} {\bibfield  {journal}
			{\bibinfo  {journal} {Phys. Rev. Res.}\ }\textbf {\bibinfo {volume} {2}},\
			\bibinfo {pages} {043184} (\bibinfo {year} {2020})}\BibitemShut {NoStop}%
		\bibitem [{\citenamefont {Gu}\ \emph {et~al.}(2017)\citenamefont {Gu},
			\citenamefont {Kockum}, \citenamefont {Miranowicz}, \citenamefont {Liu},\
			and\ \citenamefont {Nori}}]{Gu2017}%
		\BibitemOpen
		\bibfield  {author} {\bibinfo {author} {\bibfnamefont {X.}~\bibnamefont
				{Gu}}, \bibinfo {author} {\bibfnamefont {A.~F.}\ \bibnamefont {Kockum}},
			\bibinfo {author} {\bibfnamefont {A.}~\bibnamefont {Miranowicz}}, \bibinfo
			{author} {\bibfnamefont {Y.-x.}\ \bibnamefont {Liu}},\ and\ \bibinfo {author}
			{\bibfnamefont {F.}~\bibnamefont {Nori}},\ }\bibfield  {title} {\bibinfo
			{title} {Microwave photonics with superconducting quantum circuits},\ }\href
		{https://doi.org/10.1016/j.physrep.2017.10.002} {\bibfield  {journal}
			{\bibinfo  {journal} {Physics Reports}\ }\textbf {\bibinfo {volume}
				{718–719}},\ \bibinfo {pages} {1–102} (\bibinfo {year}
			{2017})}\BibitemShut {NoStop}%
		\bibitem [{\citenamefont {Kockum}\ and\ \citenamefont
			{Nori}(2019)}]{Kockum2019}%
		\BibitemOpen
		\bibfield  {author} {\bibinfo {author} {\bibfnamefont {A.~F.}\ \bibnamefont
				{Kockum}}\ and\ \bibinfo {author} {\bibfnamefont {F.}~\bibnamefont {Nori}},\
		}\bibinfo {title} {Quantum bits with {Josephson} junctions},\ in\ \href
		{https://doi.org/10.1007/978-3-030-20726-7_17} {\emph {\bibinfo {booktitle}
				{Fundamentals and Frontiers of the Josephson Effect}}}\ (\bibinfo
		{publisher} {Springer International Publishing},\ \bibinfo {year} {2019})\
		p.\ \bibinfo {pages} {703–741}\BibitemShut {NoStop}%
		\bibitem [{\citenamefont {Terradas-Brians\'o}\ \emph
			{et~al.}(2022)\citenamefont {Terradas-Brians\'o}, \citenamefont
			{Gonz\'alez-Guti\'errez}, \citenamefont {Nori}, \citenamefont
			{Mart\'{\i}n-Moreno},\ and\ \citenamefont {Zueco}}]{PhysRevA.106.063717}%
		\BibitemOpen
		\bibfield  {author} {\bibinfo {author} {\bibfnamefont {S.}~\bibnamefont
				{Terradas-Brians\'o}}, \bibinfo {author} {\bibfnamefont {C.~A.}\ \bibnamefont
				{Gonz\'alez-Guti\'errez}}, \bibinfo {author} {\bibfnamefont {F.}~\bibnamefont
				{Nori}}, \bibinfo {author} {\bibfnamefont {L.}~\bibnamefont
				{Mart\'{\i}n-Moreno}},\ and\ \bibinfo {author} {\bibfnamefont
				{D.}~\bibnamefont {Zueco}},\ }\bibfield  {title} {\bibinfo {title}
			{Ultrastrong waveguide {QED} with giant atoms},\ }\href
		{https://doi.org/10.1103/PhysRevA.106.063717} {\bibfield  {journal} {\bibinfo
				{journal} {Phys. Rev. A}\ }\textbf {\bibinfo {volume} {106}},\ \bibinfo
			{pages} {063717} (\bibinfo {year} {2022})}\BibitemShut {NoStop}%
		\bibitem [{\citenamefont {Kockum}\ \emph {et~al.}(2018)\citenamefont {Kockum},
			\citenamefont {Johansson},\ and\ \citenamefont
			{Nori}}]{PhysRevLett.120.140404}%
		\BibitemOpen
		\bibfield  {author} {\bibinfo {author} {\bibfnamefont {A.~F.}\ \bibnamefont
				{Kockum}}, \bibinfo {author} {\bibfnamefont {G.}~\bibnamefont {Johansson}},\
			and\ \bibinfo {author} {\bibfnamefont {F.}~\bibnamefont {Nori}},\ }\bibfield
		{title} {\bibinfo {title} {Decoherence-free interaction between giant atoms
				in waveguide quantum electrodynamics},\ }\href
		{https://doi.org/10.1103/PhysRevLett.120.140404} {\bibfield  {journal}
			{\bibinfo  {journal} {Phys. Rev. Lett.}\ }\textbf {\bibinfo {volume} {120}},\
			\bibinfo {pages} {140404} (\bibinfo {year} {2018})}\BibitemShut {NoStop}%
		\bibitem [{\citenamefont {Kannan}\ \emph {et~al.}(2020)\citenamefont {Kannan},
			\citenamefont {Ruckriegel}, \citenamefont {Campbell}, \citenamefont
			{Frisk~Kockum}, \citenamefont {Braum\"{u}ller}, \citenamefont {Kim},
			\citenamefont {Kjaergaard}, \citenamefont {Krantz}, \citenamefont {Melville},
			\citenamefont {Niedzielski}, \citenamefont {Veps\"{a}l\"{a}inen},
			\citenamefont {Winik}, \citenamefont {Yoder}, \citenamefont {Nori},
			\citenamefont {Orlando}, \citenamefont {Gustavsson},\ and\ \citenamefont
			{Oliver}}]{Kannan2020}%
		\BibitemOpen
		\bibfield  {author} {\bibinfo {author} {\bibfnamefont {B.}~\bibnamefont
				{Kannan}}, \bibinfo {author} {\bibfnamefont {M.~J.}\ \bibnamefont
				{Ruckriegel}}, \bibinfo {author} {\bibfnamefont {D.~L.}\ \bibnamefont
				{Campbell}}, \bibinfo {author} {\bibfnamefont {A.}~\bibnamefont
				{Frisk~Kockum}}, \bibinfo {author} {\bibfnamefont {J.}~\bibnamefont
				{Braum\"{u}ller}}, \bibinfo {author} {\bibfnamefont {D.~K.}\ \bibnamefont
				{Kim}}, \bibinfo {author} {\bibfnamefont {M.}~\bibnamefont {Kjaergaard}},
			\bibinfo {author} {\bibfnamefont {P.}~\bibnamefont {Krantz}}, \bibinfo
			{author} {\bibfnamefont {A.}~\bibnamefont {Melville}}, \bibinfo {author}
			{\bibfnamefont {B.~M.}\ \bibnamefont {Niedzielski}}, \bibinfo {author}
			{\bibfnamefont {A.}~\bibnamefont {Veps\"{a}l\"{a}inen}}, \bibinfo {author}
			{\bibfnamefont {R.}~\bibnamefont {Winik}}, \bibinfo {author} {\bibfnamefont
				{J.~L.}\ \bibnamefont {Yoder}}, \bibinfo {author} {\bibfnamefont
				{F.}~\bibnamefont {Nori}}, \bibinfo {author} {\bibfnamefont {T.~P.}\
				\bibnamefont {Orlando}}, \bibinfo {author} {\bibfnamefont {S.}~\bibnamefont
				{Gustavsson}},\ and\ \bibinfo {author} {\bibfnamefont {W.~D.}\ \bibnamefont
				{Oliver}},\ }\bibfield  {title} {\bibinfo {title} {Waveguide quantum
				electrodynamics with superconducting artificial giant atoms},\ }\href
		{https://doi.org/10.1038/s41586-020-2529-9} {\bibfield  {journal} {\bibinfo
				{journal} {Nature}\ }\textbf {\bibinfo {volume} {583}},\ \bibinfo {pages}
			{775–779} (\bibinfo {year} {2020})}\BibitemShut {NoStop}%
		\bibitem [{\citenamefont {Wang}\ \emph {et~al.}(2021)\citenamefont {Wang},
			\citenamefont {Liu}, \citenamefont {Kockum}, \citenamefont {Li},\ and\
			\citenamefont {Nori}}]{PhysRevLett.126.043602}%
		\BibitemOpen
		\bibfield  {author} {\bibinfo {author} {\bibfnamefont {X.}~\bibnamefont
				{Wang}}, \bibinfo {author} {\bibfnamefont {T.}~\bibnamefont {Liu}}, \bibinfo
			{author} {\bibfnamefont {A.~F.}\ \bibnamefont {Kockum}}, \bibinfo {author}
			{\bibfnamefont {H.-R.}\ \bibnamefont {Li}},\ and\ \bibinfo {author}
			{\bibfnamefont {F.}~\bibnamefont {Nori}},\ }\bibfield  {title} {\bibinfo
			{title} {Tunable chiral bound states with giant atoms},\ }\href
		{https://doi.org/10.1103/PhysRevLett.126.043602} {\bibfield  {journal}
			{\bibinfo  {journal} {Phys. Rev. Lett.}\ }\textbf {\bibinfo {volume} {126}},\
			\bibinfo {pages} {043602} (\bibinfo {year} {2021})}\BibitemShut {NoStop}%
		\bibitem [{\citenamefont {Wang}\ \emph {et~al.}(2022)\citenamefont {Wang},
			\citenamefont {Wang}, \citenamefont {Yao}, \citenamefont {Shen},
			\citenamefont {Wu}, \citenamefont {Qian}, \citenamefont {Li}, \citenamefont
			{Zhu},\ and\ \citenamefont {You}}]{Wang2022}%
		\BibitemOpen
		\bibfield  {author} {\bibinfo {author} {\bibfnamefont {Z.-Q.}\ \bibnamefont
				{Wang}}, \bibinfo {author} {\bibfnamefont {Y.-P.}\ \bibnamefont {Wang}},
			\bibinfo {author} {\bibfnamefont {J.}~\bibnamefont {Yao}}, \bibinfo {author}
			{\bibfnamefont {R.-C.}\ \bibnamefont {Shen}}, \bibinfo {author}
			{\bibfnamefont {W.-J.}\ \bibnamefont {Wu}}, \bibinfo {author} {\bibfnamefont
				{J.}~\bibnamefont {Qian}}, \bibinfo {author} {\bibfnamefont {J.}~\bibnamefont
				{Li}}, \bibinfo {author} {\bibfnamefont {S.-Y.}\ \bibnamefont {Zhu}},\ and\
			\bibinfo {author} {\bibfnamefont {J.~Q.}\ \bibnamefont {You}},\ }\bibfield
		{title} {\bibinfo {title} {Giant spin ensembles in waveguide magnonics},\
		}\href {http://dx.doi.org/10.1038/s41467-022-35174-9} {\bibfield  {journal}
			{\bibinfo  {journal} {Nat. Commun.}\ }\textbf {\bibinfo {volume} {13}}
			(\bibinfo {year} {2022})}\BibitemShut {NoStop}%
		\bibitem [{\citenamefont {Walls}\ and\ \citenamefont
			{Millburn}(2008)}]{walls2008optics}%
		\BibitemOpen
		\bibfield  {author} {\bibinfo {author} {\bibfnamefont {D.~F.}\ \bibnamefont
				{Walls}}\ and\ \bibinfo {author} {\bibfnamefont {G.~J.}\ \bibnamefont
				{Millburn}},\ }\href@noop {} {\emph {\bibinfo {title} {Quantum Optics}}},\
		\bibinfo {edition} {2nd}\ ed.\ (\bibinfo  {publisher} {Springer},\ \bibinfo
		{address} {Berlin},\ \bibinfo {year} {2008})\BibitemShut {NoStop}%
		\bibitem [{\citenamefont {Agarwal}(2012)}]{Agarwal2012}%
		\BibitemOpen
		\bibfield  {author} {\bibinfo {author} {\bibfnamefont {G.~S.}\ \bibnamefont
				{Agarwal}},\ }\href {https://doi.org/10.1017/cbo9781139035170} {\emph
			{\bibinfo {title} {Quantum Optics}}}\ (\bibinfo  {publisher} {Cambridge
			University Press},\ \bibinfo {year} {2012})\BibitemShut {NoStop}%
		\bibitem [{\citenamefont {Scully}\ and\ \citenamefont
			{Zubairy}(1997)}]{Marlan1997}%
		\BibitemOpen
		\bibfield  {author} {\bibinfo {author} {\bibfnamefont {M.~O.}\ \bibnamefont
				{Scully}}\ and\ \bibinfo {author} {\bibfnamefont {M.~S.}\ \bibnamefont
				{Zubairy}},\ }\href@noop {} {\emph {\bibinfo {title} {Quantum Optics}}},\
		\bibinfo {edition} {1st}\ ed.\ (\bibinfo  {publisher} {Cambridge University
			Press},\ \bibinfo {year} {1997})\BibitemShut {NoStop}%
		\bibitem [{\citenamefont {Yin}\ and\ \citenamefont
			{Liao}(2023)}]{PhysRevA.108.023728}%
		\BibitemOpen
		\bibfield  {author} {\bibinfo {author} {\bibfnamefont {X.-L.}\ \bibnamefont
				{Yin}}\ and\ \bibinfo {author} {\bibfnamefont {J.-Q.}\ \bibnamefont {Liao}},\
		}\bibfield  {title} {\bibinfo {title} {Generation of two-giant-atom
				entanglement in waveguide-{QED} systems},\ }\href
		{https://doi.org/10.1103/PhysRevA.108.023728} {\bibfield  {journal} {\bibinfo
				{journal} {Phys. Rev. A}\ }\textbf {\bibinfo {volume} {108}},\ \bibinfo
			{pages} {023728} (\bibinfo {year} {2023})}\BibitemShut {NoStop}%
		\bibitem [{\citenamefont {Shen}\ and\ \citenamefont
			{Fan}(2005)}]{PhysRevLett.95.213001}%
		\BibitemOpen
		\bibfield  {author} {\bibinfo {author} {\bibfnamefont {J.-T.}\ \bibnamefont
				{Shen}}\ and\ \bibinfo {author} {\bibfnamefont {S.}~\bibnamefont {Fan}},\
		}\bibfield  {title} {\bibinfo {title} {Coherent single photon transport in a
				one-dimensional waveguide coupled with superconducting quantum bits},\ }\href
		{https://doi.org/10.1103/PhysRevLett.95.213001} {\bibfield  {journal}
			{\bibinfo  {journal} {Phys. Rev. Lett.}\ }\textbf {\bibinfo {volume} {95}},\
			\bibinfo {pages} {213001} (\bibinfo {year} {2005})}\BibitemShut {NoStop}%
		\bibitem [{\citenamefont {Shen}\ and\ \citenamefont
			{Fan}(2009)}]{PhysRevA.79.023837}%
		\BibitemOpen
		\bibfield  {author} {\bibinfo {author} {\bibfnamefont {J.-T.}\ \bibnamefont
				{Shen}}\ and\ \bibinfo {author} {\bibfnamefont {S.}~\bibnamefont {Fan}},\
		}\bibfield  {title} {\bibinfo {title} {Theory of single-photon transport in a
				single-mode waveguide. i. coupling to a cavity containing a two-level atom},\
		}\href {https://doi.org/10.1103/PhysRevA.79.023837} {\bibfield  {journal}
			{\bibinfo  {journal} {Phys. Rev. A}\ }\textbf {\bibinfo {volume} {79}},\
			\bibinfo {pages} {023837} (\bibinfo {year} {2009})}\BibitemShut {NoStop}%
		\bibitem [{\citenamefont {Song}\ \emph {et~al.}(2024)\citenamefont {Song},
			\citenamefont {Liu}, \citenamefont {Zhou}, \citenamefont {Yang},\ and\
			\citenamefont {An}}]{PhysRevLett.132.090401}%
		\BibitemOpen
		\bibfield  {author} {\bibinfo {author} {\bibfnamefont {W.-L.}\ \bibnamefont
				{Song}}, \bibinfo {author} {\bibfnamefont {H.-B.}\ \bibnamefont {Liu}},
			\bibinfo {author} {\bibfnamefont {B.}~\bibnamefont {Zhou}}, \bibinfo {author}
			{\bibfnamefont {W.-L.}\ \bibnamefont {Yang}},\ and\ \bibinfo {author}
			{\bibfnamefont {J.-H.}\ \bibnamefont {An}},\ }\bibfield  {title} {\bibinfo
			{title} {Remote charging and degradation suppression for the quantum
				battery},\ }\href {https://doi.org/10.1103/PhysRevLett.132.090401} {\bibfield
			{journal} {\bibinfo  {journal} {Phys. Rev. Lett.}\ }\textbf {\bibinfo
				{volume} {132}},\ \bibinfo {pages} {090401} (\bibinfo {year}
			{2024})}\BibitemShut {NoStop}%
		\bibitem [{\citenamefont {Friis}\ and\ \citenamefont
			{Huber}(2018)}]{Friis2018}%
		\BibitemOpen
		\bibfield  {author} {\bibinfo {author} {\bibfnamefont {N.}~\bibnamefont
				{Friis}}\ and\ \bibinfo {author} {\bibfnamefont {M.}~\bibnamefont {Huber}},\
		}\bibfield  {title} {\bibinfo {title} {Precision and work fluctuations in
				{Gaussian} battery charging},\ }\href
		{https://doi.org/10.22331/q-2018-04-23-61} {\bibfield  {journal} {\bibinfo
				{journal} {Quantum}\ }\textbf {\bibinfo {volume} {2}},\ \bibinfo {pages} {61}
			(\bibinfo {year} {2018})}\BibitemShut {NoStop}%
		\bibitem [{YKX()}]{YKX250527}%
		\BibitemOpen
		\href@noop {} {\ }\bibinfo {note} {See {Supplemental Material} at:
			\url{https://link.aps.org/supplemental/XXXXXXX} which includes detailed
			derivations and discussions of our main results.}\BibitemShut {Stop}%
		\bibitem [{\citenamefont {Andolina}\ \emph {et~al.}(2018)\citenamefont
			{Andolina}, \citenamefont {Farina}, \citenamefont {Mari}, \citenamefont
			{Pellegrini}, \citenamefont {Giovannetti},\ and\ \citenamefont
			{Polini}}]{PhysRevB.98.205423}%
		\BibitemOpen
		\bibfield  {author} {\bibinfo {author} {\bibfnamefont {G.~M.}\ \bibnamefont
				{Andolina}}, \bibinfo {author} {\bibfnamefont {D.}~\bibnamefont {Farina}},
			\bibinfo {author} {\bibfnamefont {A.}~\bibnamefont {Mari}}, \bibinfo {author}
			{\bibfnamefont {V.}~\bibnamefont {Pellegrini}}, \bibinfo {author}
			{\bibfnamefont {V.}~\bibnamefont {Giovannetti}},\ and\ \bibinfo {author}
			{\bibfnamefont {M.}~\bibnamefont {Polini}},\ }\bibfield  {title} {\bibinfo
			{title} {Charger-mediated energy transfer in exactly solvable models for
				quantum batteries},\ }\href {https://doi.org/10.1103/PhysRevB.98.205423}
		{\bibfield  {journal} {\bibinfo  {journal} {Phys. Rev. B}\ }\textbf {\bibinfo
				{volume} {98}},\ \bibinfo {pages} {205423} (\bibinfo {year}
			{2018})}\BibitemShut {NoStop}%
		\bibitem [{\citenamefont {Zhou}\ \emph
			{et~al.}(2008{\natexlab{a}})\citenamefont {Zhou}, \citenamefont {Gong},
			\citenamefont {Liu}, \citenamefont {Sun},\ and\ \citenamefont
			{Nori}}]{PhysRevLett.101.100501}%
		\BibitemOpen
		\bibfield  {author} {\bibinfo {author} {\bibfnamefont {L.}~\bibnamefont
				{Zhou}}, \bibinfo {author} {\bibfnamefont {Z.~R.}\ \bibnamefont {Gong}},
			\bibinfo {author} {\bibfnamefont {Y.-x.}\ \bibnamefont {Liu}}, \bibinfo
			{author} {\bibfnamefont {C.~P.}\ \bibnamefont {Sun}},\ and\ \bibinfo {author}
			{\bibfnamefont {F.}~\bibnamefont {Nori}},\ }\bibfield  {title} {\bibinfo
			{title} {Controllable scattering of a single photon inside a one-dimensional
				resonator waveguide},\ }\href
		{https://doi.org/10.1103/PhysRevLett.101.100501} {\bibfield  {journal}
			{\bibinfo  {journal} {Phys. Rev. Lett.}\ }\textbf {\bibinfo {volume} {101}},\
			\bibinfo {pages} {100501} (\bibinfo {year} {2008}{\natexlab{a}})}\BibitemShut
		{NoStop}%
		\bibitem [{\citenamefont {Liao}\ \emph {et~al.}(2010)\citenamefont {Liao},
			\citenamefont {Gong}, \citenamefont {Zhou}, \citenamefont {Liu},
			\citenamefont {Sun},\ and\ \citenamefont {Nori}}]{PhysRevA.81.042304}%
		\BibitemOpen
		\bibfield  {author} {\bibinfo {author} {\bibfnamefont {J.-Q.}\ \bibnamefont
				{Liao}}, \bibinfo {author} {\bibfnamefont {Z.~R.}\ \bibnamefont {Gong}},
			\bibinfo {author} {\bibfnamefont {L.}~\bibnamefont {Zhou}}, \bibinfo {author}
			{\bibfnamefont {Y.-x.}\ \bibnamefont {Liu}}, \bibinfo {author} {\bibfnamefont
				{C.~P.}\ \bibnamefont {Sun}},\ and\ \bibinfo {author} {\bibfnamefont
				{F.}~\bibnamefont {Nori}},\ }\bibfield  {title} {\bibinfo {title}
			{Controlling the transport of single photons by tuning the frequency of
				either one or two cavities in an array of coupled cavities},\ }\href
		{https://doi.org/10.1103/PhysRevA.81.042304} {\bibfield  {journal} {\bibinfo
				{journal} {Phys. Rev. A}\ }\textbf {\bibinfo {volume} {81}},\ \bibinfo
			{pages} {042304} (\bibinfo {year} {2010})}\BibitemShut {NoStop}%
		\bibitem [{\citenamefont {Zhou}\ \emph {et~al.}(2009)\citenamefont {Zhou},
			\citenamefont {Yang}, \citenamefont {Liu}, \citenamefont {Sun},\ and\
			\citenamefont {Nori}}]{PhysRevA.80.062109}%
		\BibitemOpen
		\bibfield  {author} {\bibinfo {author} {\bibfnamefont {L.}~\bibnamefont
				{Zhou}}, \bibinfo {author} {\bibfnamefont {S.}~\bibnamefont {Yang}}, \bibinfo
			{author} {\bibfnamefont {Y.-x.}\ \bibnamefont {Liu}}, \bibinfo {author}
			{\bibfnamefont {C.~P.}\ \bibnamefont {Sun}},\ and\ \bibinfo {author}
			{\bibfnamefont {F.}~\bibnamefont {Nori}},\ }\bibfield  {title} {\bibinfo
			{title} {Quantum {Zeno} switch for single-photon coherent transport},\ }\href
		{https://doi.org/10.1103/PhysRevA.80.062109} {\bibfield  {journal} {\bibinfo
				{journal} {Phys. Rev. A}\ }\textbf {\bibinfo {volume} {80}},\ \bibinfo
			{pages} {062109} (\bibinfo {year} {2009})}\BibitemShut {NoStop}%
		\bibitem [{\citenamefont {Zhou}\ \emph
			{et~al.}(2008{\natexlab{b}})\citenamefont {Zhou}, \citenamefont {Dong},
			\citenamefont {Liu}, \citenamefont {Sun},\ and\ \citenamefont
			{Nori}}]{PhysRevA.78.063827}%
		\BibitemOpen
		\bibfield  {author} {\bibinfo {author} {\bibfnamefont {L.}~\bibnamefont
				{Zhou}}, \bibinfo {author} {\bibfnamefont {H.}~\bibnamefont {Dong}}, \bibinfo
			{author} {\bibfnamefont {Y.-x.}\ \bibnamefont {Liu}}, \bibinfo {author}
			{\bibfnamefont {C.~P.}\ \bibnamefont {Sun}},\ and\ \bibinfo {author}
			{\bibfnamefont {F.}~\bibnamefont {Nori}},\ }\bibfield  {title} {\bibinfo
			{title} {Quantum supercavity with atomic mirrors},\ }\href
		{https://doi.org/10.1103/PhysRevA.78.063827} {\bibfield  {journal} {\bibinfo
				{journal} {Phys. Rev. A}\ }\textbf {\bibinfo {volume} {78}},\ \bibinfo
			{pages} {063827} (\bibinfo {year} {2008}{\natexlab{b}})}\BibitemShut
		{NoStop}%
		\bibitem [{\citenamefont {Wang}\ and\ \citenamefont {Li}(2022)}]{W2022}%
		\BibitemOpen
		\bibfield  {author} {\bibinfo {author} {\bibfnamefont {X.}~\bibnamefont
				{Wang}}\ and\ \bibinfo {author} {\bibfnamefont {H.-R.}\ \bibnamefont {Li}},\
		}\bibfield  {title} {\bibinfo {title} {Chiral quantum network with giant
				atoms},\ }\href {https://doi.org/10.1088/2058-9565/ac6a04} {\bibfield
			{journal} {\bibinfo  {journal} {Quantum Sci. Technol.}\ }\textbf {\bibinfo
				{volume} {7}},\ \bibinfo {pages} {035007} (\bibinfo {year}
			{2022})}\BibitemShut {NoStop}%
		\bibitem [{\citenamefont {Wulschner}\ \emph {et~al.}(2016)\citenamefont
			{Wulschner}, \citenamefont {Goetz}, \citenamefont {Koessel}, \citenamefont
			{Hoffmann}, \citenamefont {Baust}, \citenamefont {Eder}, \citenamefont
			{Fischer}, \citenamefont {Haeberlein}, \citenamefont {Schwarz}, \citenamefont
			{Pernpeintner}, \citenamefont {Xie}, \citenamefont {Zhong}, \citenamefont
			{Zollitsch}, \citenamefont {Peropadre}, \citenamefont {Garcia~Ripoll},
			\citenamefont {Solano}, \citenamefont {Fedorov}, \citenamefont {Menzel},
			\citenamefont {Deppe}, \citenamefont {Marx},\ and\ \citenamefont
			{Gross}}]{Wulschner2016}%
		\BibitemOpen
		\bibfield  {author} {\bibinfo {author} {\bibfnamefont {F.}~\bibnamefont
				{Wulschner}}, \bibinfo {author} {\bibfnamefont {J.}~\bibnamefont {Goetz}},
			\bibinfo {author} {\bibfnamefont {F.~R.}\ \bibnamefont {Koessel}}, \bibinfo
			{author} {\bibfnamefont {E.}~\bibnamefont {Hoffmann}}, \bibinfo {author}
			{\bibfnamefont {A.}~\bibnamefont {Baust}}, \bibinfo {author} {\bibfnamefont
				{P.}~\bibnamefont {Eder}}, \bibinfo {author} {\bibfnamefont {M.}~\bibnamefont
				{Fischer}}, \bibinfo {author} {\bibfnamefont {M.}~\bibnamefont {Haeberlein}},
			\bibinfo {author} {\bibfnamefont {M.~J.}\ \bibnamefont {Schwarz}}, \bibinfo
			{author} {\bibfnamefont {M.}~\bibnamefont {Pernpeintner}}, \bibinfo {author}
			{\bibfnamefont {E.}~\bibnamefont {Xie}}, \bibinfo {author} {\bibfnamefont
				{L.}~\bibnamefont {Zhong}}, \bibinfo {author} {\bibfnamefont {C.~W.}\
				\bibnamefont {Zollitsch}}, \bibinfo {author} {\bibfnamefont {B.}~\bibnamefont
				{Peropadre}}, \bibinfo {author} {\bibfnamefont {J.-J.}\ \bibnamefont
				{Garcia~Ripoll}}, \bibinfo {author} {\bibfnamefont {E.}~\bibnamefont
				{Solano}}, \bibinfo {author} {\bibfnamefont {K.~G.}\ \bibnamefont {Fedorov}},
			\bibinfo {author} {\bibfnamefont {E.~P.}\ \bibnamefont {Menzel}}, \bibinfo
			{author} {\bibfnamefont {F.}~\bibnamefont {Deppe}}, \bibinfo {author}
			{\bibfnamefont {A.}~\bibnamefont {Marx}},\ and\ \bibinfo {author}
			{\bibfnamefont {R.}~\bibnamefont {Gross}},\ }\bibfield  {title} {\bibinfo
			{title} {Tunable coupling of transmission-line microwave resonators mediated
				by an {RF SQUID}},\ }\href
		{http://dx.doi.org/10.1140/epjqt/s40507-016-0048-2} {\bibfield  {journal}
			{\bibinfo  {journal} {EPJ Quantum Technol.}\ }\textbf {\bibinfo {volume} {3}}
			(\bibinfo {year} {2016})}\BibitemShut {NoStop}%
		\bibitem [{\citenamefont {Geller}\ \emph {et~al.}(2015)\citenamefont {Geller},
			\citenamefont {Donate}, \citenamefont {Chen}, \citenamefont {Fang},
			\citenamefont {Leung}, \citenamefont {Neill}, \citenamefont {Roushan},\ and\
			\citenamefont {Martinis}}]{PhysRevA.92.012320}%
		\BibitemOpen
		\bibfield  {author} {\bibinfo {author} {\bibfnamefont {M.~R.}\ \bibnamefont
				{Geller}}, \bibinfo {author} {\bibfnamefont {E.}~\bibnamefont {Donate}},
			\bibinfo {author} {\bibfnamefont {Y.}~\bibnamefont {Chen}}, \bibinfo {author}
			{\bibfnamefont {M.~T.}\ \bibnamefont {Fang}}, \bibinfo {author}
			{\bibfnamefont {N.}~\bibnamefont {Leung}}, \bibinfo {author} {\bibfnamefont
				{C.}~\bibnamefont {Neill}}, \bibinfo {author} {\bibfnamefont
				{P.}~\bibnamefont {Roushan}},\ and\ \bibinfo {author} {\bibfnamefont {J.~M.}\
				\bibnamefont {Martinis}},\ }\bibfield  {title} {\bibinfo {title} {Tunable
				coupler for superconducting {X-mon} qubits: Perturbative nonlinear model},\
		}\href {https://doi.org/10.1103/PhysRevA.92.012320} {\bibfield  {journal}
			{\bibinfo  {journal} {Phys. Rev. A}\ }\textbf {\bibinfo {volume} {92}},\
			\bibinfo {pages} {012320} (\bibinfo {year} {2015})}\BibitemShut {NoStop}%
		\bibitem [{\citenamefont {Stannigel}\ \emph {et~al.}(2011)\citenamefont
			{Stannigel}, \citenamefont {Rabl}, \citenamefont {S\o{}rensen}, \citenamefont
			{Lukin},\ and\ \citenamefont {Zoller}}]{PhysRevA.84.042341}%
		\BibitemOpen
		\bibfield  {author} {\bibinfo {author} {\bibfnamefont {K.}~\bibnamefont
				{Stannigel}}, \bibinfo {author} {\bibfnamefont {P.}~\bibnamefont {Rabl}},
			\bibinfo {author} {\bibfnamefont {A.~S.}\ \bibnamefont {S\o{}rensen}},
			\bibinfo {author} {\bibfnamefont {M.~D.}\ \bibnamefont {Lukin}},\ and\
			\bibinfo {author} {\bibfnamefont {P.}~\bibnamefont {Zoller}},\ }\bibfield
		{title} {\bibinfo {title} {Optomechanical transducers for quantum-information
				processing},\ }\href {https://doi.org/10.1103/PhysRevA.84.042341} {\bibfield
			{journal} {\bibinfo  {journal} {Phys. Rev. A}\ }\textbf {\bibinfo {volume}
				{84}},\ \bibinfo {pages} {042341} (\bibinfo {year} {2011})}\BibitemShut
		{NoStop}%
		\bibitem [{\citenamefont {Vermersch}\ \emph {et~al.}(2017)\citenamefont
			{Vermersch}, \citenamefont {Guimond}, \citenamefont {Pichler},\ and\
			\citenamefont {Zoller}}]{PhysRevLett.118.133601}%
		\BibitemOpen
		\bibfield  {author} {\bibinfo {author} {\bibfnamefont {B.}~\bibnamefont
				{Vermersch}}, \bibinfo {author} {\bibfnamefont {P.-O.}\ \bibnamefont
				{Guimond}}, \bibinfo {author} {\bibfnamefont {H.}~\bibnamefont {Pichler}},\
			and\ \bibinfo {author} {\bibfnamefont {P.}~\bibnamefont {Zoller}},\
		}\bibfield  {title} {\bibinfo {title} {Quantum state transfer via noisy
				photonic and phononic waveguides},\ }\href
		{https://doi.org/10.1103/PhysRevLett.118.133601} {\bibfield  {journal}
			{\bibinfo  {journal} {Phys. Rev. Lett.}\ }\textbf {\bibinfo {volume} {118}},\
			\bibinfo {pages} {133601} (\bibinfo {year} {2017})}\BibitemShut {NoStop}%
		\bibitem [{\citenamefont {Andersson}\ \emph {et~al.}(2019)\citenamefont
			{Andersson}, \citenamefont {Suri}, \citenamefont {Guo}, \citenamefont
			{Aref},\ and\ \citenamefont {Delsing}}]{Andersson2019}%
		\BibitemOpen
		\bibfield  {author} {\bibinfo {author} {\bibfnamefont {G.}~\bibnamefont
				{Andersson}}, \bibinfo {author} {\bibfnamefont {B.}~\bibnamefont {Suri}},
			\bibinfo {author} {\bibfnamefont {L.}~\bibnamefont {Guo}}, \bibinfo {author}
			{\bibfnamefont {T.}~\bibnamefont {Aref}},\ and\ \bibinfo {author}
			{\bibfnamefont {P.}~\bibnamefont {Delsing}},\ }\bibfield  {title} {\bibinfo
			{title} {Non-exponential decay of a giant artificial atom},\ }\href
		{https://doi.org/10.1038/s41567-019-0605-6} {\bibfield  {journal} {\bibinfo
				{journal} {Nat. Phys.}\ }\textbf {\bibinfo {volume} {15}},\ \bibinfo {pages}
			{1123–1127} (\bibinfo {year} {2019})}\BibitemShut {NoStop}%
		\bibitem [{\citenamefont {Gustafsson}\ \emph {et~al.}(2014)\citenamefont
			{Gustafsson}, \citenamefont {Aref}, \citenamefont {Kockum}, \citenamefont
			{Ekstr\"{o}m}, \citenamefont {Johansson},\ and\ \citenamefont
			{Delsing}}]{Gustafsson2014}%
		\BibitemOpen
		\bibfield  {author} {\bibinfo {author} {\bibfnamefont {M.~V.}\ \bibnamefont
				{Gustafsson}}, \bibinfo {author} {\bibfnamefont {T.}~\bibnamefont {Aref}},
			\bibinfo {author} {\bibfnamefont {A.~F.}\ \bibnamefont {Kockum}}, \bibinfo
			{author} {\bibfnamefont {M.~K.}\ \bibnamefont {Ekstr\"{o}m}}, \bibinfo
			{author} {\bibfnamefont {G.}~\bibnamefont {Johansson}},\ and\ \bibinfo
			{author} {\bibfnamefont {P.}~\bibnamefont {Delsing}},\ }\bibfield  {title}
		{\bibinfo {title} {Propagating phonons coupled to an artificial atom},\
		}\href {https://doi.org/10.1126/science.1257219} {\bibfield  {journal}
			{\bibinfo  {journal} {Science}\ }\textbf {\bibinfo {volume} {346}},\ \bibinfo
			{pages} {207–211} (\bibinfo {year} {2014})}\BibitemShut {NoStop}%
		\bibitem [{\citenamefont {Manenti}\ \emph {et~al.}(2017)\citenamefont
			{Manenti}, \citenamefont {Kockum}, \citenamefont {Patterson}, \citenamefont
			{Behrle}, \citenamefont {Rahamim}, \citenamefont {Tancredi}, \citenamefont
			{Nori},\ and\ \citenamefont {Leek}}]{Manenti2017}%
		\BibitemOpen
		\bibfield  {author} {\bibinfo {author} {\bibfnamefont {R.}~\bibnamefont
				{Manenti}}, \bibinfo {author} {\bibfnamefont {A.~F.}\ \bibnamefont {Kockum}},
			\bibinfo {author} {\bibfnamefont {A.}~\bibnamefont {Patterson}}, \bibinfo
			{author} {\bibfnamefont {T.}~\bibnamefont {Behrle}}, \bibinfo {author}
			{\bibfnamefont {J.}~\bibnamefont {Rahamim}}, \bibinfo {author} {\bibfnamefont
				{G.}~\bibnamefont {Tancredi}}, \bibinfo {author} {\bibfnamefont
				{F.}~\bibnamefont {Nori}},\ and\ \bibinfo {author} {\bibfnamefont {P.~J.}\
				\bibnamefont {Leek}},\ }\bibfield  {title} {\bibinfo {title} {Circuit quantum
				acoustodynamics with surface acoustic waves},\ }\href
		{http://dx.doi.org/10.1038/s41467-017-01063-9} {\bibfield  {journal}
			{\bibinfo  {journal} {Nat. Commun.}\ }\textbf {\bibinfo {volume} {8}}
			(\bibinfo {year} {2017})}\BibitemShut {NoStop}%
		\bibitem [{\citenamefont {Blais}\ \emph {et~al.}(2004)\citenamefont {Blais},
			\citenamefont {Huang}, \citenamefont {Wallraff}, \citenamefont {Girvin},\
			and\ \citenamefont {Schoelkopf}}]{PhysRevA.69.062320}%
		\BibitemOpen
		\bibfield  {author} {\bibinfo {author} {\bibfnamefont {A.}~\bibnamefont
				{Blais}}, \bibinfo {author} {\bibfnamefont {R.-S.}\ \bibnamefont {Huang}},
			\bibinfo {author} {\bibfnamefont {A.}~\bibnamefont {Wallraff}}, \bibinfo
			{author} {\bibfnamefont {S.~M.}\ \bibnamefont {Girvin}},\ and\ \bibinfo
			{author} {\bibfnamefont {R.~J.}\ \bibnamefont {Schoelkopf}},\ }\bibfield
		{title} {\bibinfo {title} {Cavity quantum electrodynamics for superconducting
				electrical circuits: An architecture for quantum computation},\ }\href
		{https://doi.org/10.1103/PhysRevA.69.062320} {\bibfield  {journal} {\bibinfo
				{journal} {Phys. Rev. A}\ }\textbf {\bibinfo {volume} {69}},\ \bibinfo
			{pages} {062320} (\bibinfo {year} {2004})}\BibitemShut {NoStop}%
	\end{thebibliography}
	%

\end{document}